\documentclass{jfm}

\usepackage{graphicx}
\usepackage{caption}
\usepackage{subcaption}
\usepackage{hyperref} 
\usepackage{natbib,amsmath}
\usepackage{mathtools}
\usepackage[dvipsnames]{xcolor}

\ifCUPmtlplainloaded \else
  \checkfont{eurm10}
  \iffontfound
    \IfFileExists{upmath.sty}
      {\typeout{^^JFound AMS Euler Roman fonts on the system,
                   using the 'upmath' package.^^J}%
       \usepackage{upmath}}
      {\typeout{^^JFound AMS Euler Roman fonts on the system, but you
                   dont seem to have the}%
       \typeout{'upmath' package installed. JFM.cls can take advantage
                 of these fonts,^^Jif you use 'upmath' package.^^J}%
      }
  \else
  \fi
\fi

\ifCUPmtlplainloaded \else
  \checkfont{msam10}
       \iffontfound
    \IfFileExists{amssymb.sty}
      {\typeout{^^JFound AMS Symbol fonts on the system, using the
                'amssymb' package.^^J}%
       \usepackage{amssymb}%

      }{}
  \fi
\fi

\ifCUPmtlplainloaded \else
  \IfFileExists{amsbsy.sty}
    {\typeout{^^JFound the 'amsbsy' package on the system, using it.^^J}%
     \usepackage{amsbsy}}
    {\providecommand\boldsymbol[1]{\mbox{\boldmath $##1$}}}
\fi

\newsavebox{\astrutbox}
\sbox{\astrutbox}{\rule[-5pt]{0pt}{20pt}}

  \newcommand{\rmd}{{\rm d}}
\newcommand{\bx}{{ \boldsymbol{x} }}
\newcommand{\bA}{{ \boldsymbol{A} }}
\newcommand{\bX}{{ \boldsymbol{X} }}
\newcommand{\tbA}{{ \widetilde{\boldsymbol{A}} }}
\newcommand{\tbD}{{ \widetilde{\boldsymbol{D}} }}

\newcommand{\bP}{{\mathbb{P}}}
\newcommand{\bE}{{\mathbb{E}}}
\newcommand{\bW}{{\wt{\mathbf{W}}}}
\newcommand{\bu}{{ \boldsymbol{u}}}

\newcommand{\bom}{{\mbox{\boldmath $\omega$}}}
\newcommand{\hd}{\hat{\rmd}}

\newcommand{\wt}{\widetilde}
\newcommand{\ta}{{\tilde{a}}}
\newcommand{\tb}{{\tilde{b}}}
\newcommand{\tc}{{\tilde{c}}}

\newcommand{\tsigma}{{\tilde{\sigma}}}

\newcommand{\ba}{{\boldsymbol{a}}}
\newcommand{\bb}{{\boldsymbol{b}}}

\newcommand{\bm}{{\boldsymbol{{\rm p}}}}
\newcommand{\bn}{\hat{{\boldsymbol{n}}}}

\newcommand{\cL}{{\mathcal{L}}}

\newcommand{\grad}{{\mbox{\boldmath $\nabla$}}}
\newcommand{\btimes}{{\mbox{\boldmath $\times$}}}
\newcommand{\bzed}{{\bold 0}}
\newcommand{\oo}{{\mbox{$\bar{{\rm o}}$}}}

\newcommand{\bsigma}{{\mbox{\boldmath $\sigma$}}}
\newcommand{\bSigma}{{\mbox{\boldmath $\Sigma$}}}

\newcommand{\be}{\begin{equation}}
\newcommand{\ee}{\end{equation}} 
\newcommand{\lb}{\label}

\def\Xint#1{\mathchoice
   {\XXint\displaystyle\textstyle{#1}}%
   {\XXint\textstyle\scriptstyle{#1}}%
   {\XXint\scriptstyle\scriptscriptstyle{#1}}%
   {\XXint\scriptscriptstyle\scriptscriptstyle{#1}}%
   \!\int}
\def\XXint#1#2#3{{\setbox0=\hbox{$#1{#2#3}{\int}$}
     \vcenter{\hbox{$#2#3$}}\kern-.5\wd0}}

\def\dashint{\Xint-}

\title[Stochastic Lagrangian Dynamics]{Stochastic Lagrangian Dynamics of Vorticity.\\
I. General Theory}

\author[G. L. Eyink et al.]%
{G. L. Eyink$^{1,2,3}$, A. Gupta$^{1}$, and T. Zaki$^{3}$}

\affiliation{$^1$Department of Applied Mathematics \& Statistics, The Johns Hopkins University, Baltimore, MD 21218, USA\\[\affilskip]
$^2$Department of Physics \& Astronomy, The Johns Hopkins University, Baltimore, MD 21218, USA\\[\affilskip]
$^3$Department of Mechanical Engineering, The Johns Hopkins University, Baltimore, MD 21218, USA}

\pubyear{2020}
\volume{???}
\pagerange{???}
\date{?; revised ?; accepted ?. - To be entered by editorial office}
\begin{document}

\maketitle

\begin{abstract}
Prior mathematical work of \cite{ConstantinIyer08,ConstantinIyer11} has shown that incompressible 
Navier-Stokes solutions possess infinitely-many stochastic Lagrangian conservation laws for 
vorticity, backward in time, which generalize the invariants of \cite{cauchy1815theorie} for smooth 
Euler solutions. We simplify this theory for the case of wall-bounded flows by appealing to the  
\cite{kuzmin1983ideal}-\cite{oseledets1989new} representation of Navier-Stokes dynamics,  
in terms of the vortex-momentum density associated to a continuous distribution of infinitesimal 
vortex rings. The Constantin-Iyer theory provides an exact representation for vorticity at any 
interior point as an average over stochastic vorticity contributions transported from the wall. 
We discuss relations of this Lagrangian formulation with the Eulerian theory of 
\cite{lighthill1963boundary}-\cite{morton1984generation} for vorticity generation at solid walls,
and also with a statistical result of \cite{taylor1932transport}-\cite{huggins1994vortex}, which connects 
dissipative drag with organized cross-stream motion of vorticity and which is closely analogous to the 
``Josephson-Anderson relation'' for quantum superfluids. We elaborate a Monte Carlo numerical 
Lagrangian scheme to calculate the stochastic Cauchy invariants and their statistics, given the 
Eulerian space-time velocity field. The method is validated using an online database of a turbulent 
channel-flow simulation \citep{graham2016web}, where  conservation of the mean Cauchy invariant 
is verified for two selected buffer-layer events corresponding to an ``ejection'' and a ``sweep''. 
The variances of the stochastic Cauchy invariants grow exponentially backward in time, however, 
confirming earlier observations of Lagrangian chaos in channel-flow turbulence.   
\end{abstract}






\section{Introduction}

Vorticity in inviscid fluids governed by incompressible Euler equations has remarkable geometric 
and Lagrangian properties that were derived in classic works of \cite{cauchy1815theorie}, 
\cite{helmholtz1858uber}, \cite{weber1868ueber}, \cite{kelvin1868vi} 
and others. In a Hamiltonian formulation of the ideal fluid equations, these properties arise from  
an infinite-dimensional particle-relabelling symmetry of the action function \citep{salmon1988hamiltonian}
and they are intimately related to the modern geometric view of the Euler equations as governing 
geodesic flow on the space {\it SDiff} of volume-preserving transformations \citep{arnold1966geometrie,besse2017geometric}. 
These Lagrangian properties of vorticity have often been invoked in theories of turbulent energy dissipation,
especially by \cite{taylor1937statistical,taylor1938production} and \cite{taylor1937mechanism}. In these 
papers Taylor suggested that vortex lines which are materially advected by a turbulent flow should be 
lengthened by random stretching and mean-square vorticity thus enhanced by conservation of 
circulations.  Taylor admitted, however, that vortex lines move as material lines and circulations 
on material loops are conserved only for ideal smooth Euler flows and these properties may be substantially 
modified by fluid viscosity. Subsequent works have confirmed that statistical and dynamical properties of 
vortex lines and material lines are quite distinct in turbulent flows. For example, 
\cite{luthi2005lagrangian,guala2005evolution,guala2006stretching} have found in an experimental  study of nearly homogeneous, 
isotropic turbulence that vortex stretching rates are smaller than rates of material-line stretching when both 
are measured along the trajectories of neutrally-buoyant polystyrene particles. A closely related numerical 
study of isotropic turbulence by \cite{johnson2016large} has shown that these differences in stretching 
rates extend even to the rare fluctuations described by large-deviations theory. \cite{johnson2017analysis} have 
furthermore shown that large-deviation statistics of vortex-stretching rates and material line-stretching rates are 
distinct also in turbulent channel flow. It might be inferred from these negative findings that the remarkable 
Lagrangian properties enjoyed by vorticity for smooth Euler flows have no relevance for physical turbulent flows, 
even at very high Reynolds numbers.   
 
Recently, however, \cite{ConstantinIyer08,ConstantinIyer11} have shown that the deep geometric 
and Lagrangian properties of vorticity for Euler dynamics fully extend to viscous Navier-Stokes solutions 
within a stochastic framework.  In the setting of their theorems, a random white-noise is added to the 
evolution equations for Lagrangian fluid particles and this noise averaged over to represent mathematically 
the viscous diffusion of momentum. \cite{ConstantinIyer08,ConstantinIyer11} showed that every 
Navier-Stokes solution satisfies stochastic generalizations of the Weber formula \citep{weber1868ueber},
the conservation of circulations \citep{kelvin1868vi}, the Cauchy formula and conserved Cauchy 
invariants \citep{cauchy1815theorie}, with all of these results reducing in the inviscid limit to those satisfied 
by smooth Euler solutions. \cite{eyink2010stochastic} has shown that these properties follow from particle-relabelling symmetry 
in a stochastic least-action formulation of incompressible Navier-Stokes  and \cite{rezakhanlou2016stochastically}    
has discussed the Constantin-Iyer results in the setting of stochastic symplectic flows in phase-space.
It should be noted that the same stochastic Lagrangian representation of vorticity as \cite{ConstantinIyer08} 
was obtained even earlier by \cite{rapoport2002random}, section 11, as the corollary of a more general result for 
Navier-Stokes equations on a compact Riemannian manifold without boundary.  This level of generality is not required for
most applications in fluid mechanics and, unfortunately, obscures the simplicity of the results for 
Navier-Stokes flows in flat Euclidean space. 

The established Lagrangian properties of vorticity hold both for boundary-free Navier-Stokes flows \citep{ConstantinIyer08} 
and for wall-bounded flows with stick b.c. on the fluid velocity \citep{ConstantinIyer11}.  In the latter case, however,
\cite{ConstantinIyer11} were able to show only existence of suitable boundary conditions in their Lagrangian
formulation, without an explicit, concrete construction. We solve that problem here by exploiting the 
Kuz'min-Oseledets formulation of the incompressible Navier-Stokes equation in terms of the vortex-momentum 
density \citep{kuzmin1983ideal,oseledets1989new}. We recall that vortex momentum is the total impulse 
required to set  a compact vortex into motion 
(\cite{batchelor2000introduction}, section 7.2) and its density can be 
interpreted physically as the vortex-momentum per volume when the fluid velocity field is represented by a 
continuous distribution of infinitesimal vortex-rings.  The theory of \cite{ConstantinIyer08,ConstantinIyer11} can be
understood as a stochastic representation of viscosity in the Kuz'min-Oseledets formulation, which we show
extends readily to wall-bounded flows. In this setting the \cite{ConstantinIyer11} theory provides an infinite set of 
stochastic Cauchy invariants for Navier-Stokes solutions, which permit an exact representation of the vorticity 
at any interior point in terms of vorticity at the wall. We further discuss some relations of \cite{ConstantinIyer11} 
with the standard Eulerian theory of vorticity generation at solid boundaries due to \cite{lighthill1963boundary} and 
\cite{morton1984generation}, which relates inviscid generation of vorticity to tangential pressure gradients at the wall.   
Wall-bounded turbulence driven by an imposed pressure-gradient or 
freestream velocity differs from homogeneous, isotropic turbulence 
in that energy dissipation requires not just random stretching of vortex lines but in fact an organized, coherent 
transport of vorticity. Indeed it follows from the Lighthill-Morton theory that a mean pressure-drop down a pipe
or channel implies the production of spanwise vorticity and a mean viscous flux of such vorticity away from the wall.  
This exact relation has been previously exploited in developing turbulent drag-control methods aimed 
at modifying the boundary vorticity flux \citep{koumoutsakos1999vorticity,zhao2004turbulent}. 

It is in fact a paradigm in the study of quantum fluids that effective drag in an otherwise non-dissipative superfluid 
is associated to cross-stream motion of quantized vortex lines by the so-called ``Josepson-Anderson relation'';
see \cite{josephson1962possible,anderson1966considerations}, and \cite{packard1998role,varoquaux2015anderson} 
for reviews. Such a result was already anticipated for classical turbulent pipe-flow by \cite{taylor1932transport} and more 
systematically discussed in that context by \cite{huggins1970energy,huggins1994vortex} and \cite{eyink2008turbulent}. 
The physical basis of the relation is very general and follows just from the local conservation of momentum,
rewritten as  
\be \partial_t u_i =\frac{1}{2}\epsilon_{ijk} \Sigma_{jk} - \partial_i h \lb{eq1}  \ee
for suitable anti-symmetric matrix $\bSigma$ and potential $h.$ For example, the incompressible Navier-Stokes 
equation when rewritten in this manner has these quantities given by 
\be \Sigma_{ij} = u_i\omega_j - u_j\omega_i + \nu\left(\frac{\partial\omega_i}{\partial x_j}-
\frac{\partial\omega_j}{\partial x_i}\right), \quad h = p +\frac{1}{2}|\bu|^2.  \lb{eq2} \ee
Other fluid systems will have different contributions here, e.g. a body-force ${\bf f}$ from the stress of a polymer 
additive will contribute a term $\Sigma_{ij}=\epsilon_{ijk}f_k$ from the Magnus effect.  Taking the curl of 
\eqref{eq1} in general leads to the equation for local conservation of vorticity 
\be  \partial_t \omega_j + \partial_i \Sigma_{ij} =0 \lb{eq3} \ee  
where $\Sigma_{ij}$ gives the flux of $j$th component of vorticity in the $i$th coordinate direction. 
For Navier-Stokes, \eqref{eq3} is equivalent to the equation of \cite{helmholtz1858uber}, with contributions 
to the vorticity flux in \eqref{eq2} arising from nonlinear advection, vortex-stretching and viscous 
diffusion, respectively. As pointed out by \cite{taylor1932transport}  for 2D turbulent pipe flow and by 
\cite{huggins1994vortex}  for more general statistically steady flows, the average of \eqref{eq1} implies 
a pointwise relation  between mean vorticity transport and the mean gradient of (generalized) pressure:  
\be  \overline{\Sigma}_{ij} = \epsilon_{ijk} \partial_k \overline{\left( p+\frac{1}{2}|\bu|^2\right)}  \lb{eq4} \ee 
with $\partial_i \overline{\Sigma}_{ij}=0.$ There is an inward flux of spanwise vorticity not only at the wall,
as predicted by the Lighthill-Morton theory, but also in the bulk of the flow, intrinsically linked 
to a downstream drop in pressure or in kinetic energy density. Such vorticity flux occurs not only 
in pipes and channels but also in many other cases, such as turbulent free shear layers and wakes
\citep{brown2012turbulent}. The \cite{ConstantinIyer11} theory provides a Lagrangian description of this 
interior vorticity transport in terms of the stochastic motion and deformation of vorticity vector elements. 

The detailed contents of this paper are outlined as follows. In section \ref{sec:NS} we present 
the stochastic Lagrangian formulation of Navier-Stokes, reviewing the Kuz'min-Oseledets 
formulation (\ref{sec:KO}), explaining the Constantin-Iyer theory within this framework 
(\ref{sec:CI}), and discussing the relation with the theories of Lighthill-Morton and Taylor-Huggins   
(\ref{sec:LM}). In section \ref{sec:num}  we present and evaluate a numerical implementation,
elaborating our Monte Carlo Lagrangian scheme (\ref{sec:time-int}), 
describing the channel-flow database used in the study (\ref{sec:JHTDB}),
and validating the numerical method (\ref{sec:valid}).
The conclusion section \ref{sec:concl} discusses some open issues and prospects
for future work, and an Appendix \ref{sec:interp} provides some technical details on 
stochastic interpolation. Additional information is provided as Supplemental
Materials (SM). In the following paper of 
\cite{eyink2020stochasticII} (hereafter referenced as Paper II in the text)  
we exploit the methods of the present work to study in detail the process of 
vortex-lifting in the buffer-layer of turbulent channel-flow. 

\section{Stochastic Lagrangian Formulation of the Navier-Stokes Equation}\lb{sec:NS} 

The stochastic Lagrangian representation of incompressible Navier-Stokes solutions due to 
\cite{ConstantinIyer08,ConstantinIyer11} 
was developed by them using the 
\cite{weber1868ueber} formula.  
We believe that their work is better understood, however, using the closely related 
``vortex-momentum'' formalism of 
\cite{kuzmin1983ideal} and \cite{oseledets1989new}.  
Although originally 
derived for incompressible Euler and Navier-Stokes equations in all of Euclidean space, 
the Kuz'min-Oseledets formalism carries over very naturally to wall-bounded flows. We 
review the latter theory, first for incompressible Euler equations and then for Navier-Stokes 
equations at non-vanishing viscosity, both suitably generalized to flows with solid walls. 
Next, we explain the stochastic representations of 
\cite{ConstantinIyer08,ConstantinIyer11}
in this framework, 
simplifying in particular their own discussion of boundary conditions. As we explain, their theory yields 
infinitely many stochastic Lagrangian conservation laws for incompressible Navier-Stokes solutions,  
which generalize the vorticity invariants of 
\cite{cauchy1815theorie} 
for Euler solutions. These stochastic 
Cauchy invariants provide a precise mathematical tool with which to trace the evolution of interior 
vorticity back to its origins at the solid wall. Finally, we shall briefly compare this framework with 
analytic theories for vorticity-generation at the flow boundary, in particular the theory of 
\cite{lighthill1963boundary} and \cite{morton1984generation} 
and related work of 
\cite{taylor1932transport} and \cite{huggins1970energy,huggins1994vortex}.

\subsection{Kuz'min-Oseledets Representation of Euler and Navier-Stokes}\lb{sec:KO} 

We begin with a few mathematical preliminaries. The Kuz'min-Oseledets theory employs essentially 
the Leray-Hodge projection $\bP$ operator on vector fields, which projects to the solenoidal 
component (\cite{boyer2012mathematical}, section 3.3). In flow domains $\Omega$ with a non-empty 
boundary $\partial\Omega\neq \emptyset$, this projection operator is defined on a vector field $\bm$ by 
\be \bP\bm = \bm-\grad_x\phi,  \lb{eq2-1} \ee
where $\phi$ is the velocity potential solving the Neumann boundary-value problem:  
\be \triangle_x\phi =\grad_x\cdot\bm, \quad \bx\in \Omega; \qquad  \frac{\partial\phi}{\partial n} =\bn\cdot\bm,
\quad \bx\in \partial\Omega. \lb{eq2-2} \ee 
Here $\bn$ is the outward-pointing unit normal vector at a point on the smooth boundary $\partial\Omega.$
This projection yields the Hodge decomposition $\bm=\bu+\grad_x\phi,$ where $\bu=\bP\bm$ satisfies $\grad_x\cdot\bu=0$ 
on $\Omega$ and $\bn\cdot\bu=0$ on $\partial\Omega.$ All of the interior vorticity of the flow 
arises then obviously from the solenoidal component $\bu.$ On a matter of notation, we shall 
use in our discussion below the convention of ``dyadic products" of vectors, with $\ba\bb$ 
representing the tensor product, usually denoted by $\ba\otimes\bb$ in the mathematical literature, so 
that $(\ba\bb)_{ij}=a_ib_j.$ This convention applies in particular to Jacobian derivatives $\grad_x\ba,$ 
so that $(\grad_x\ba)_{ij}=\partial_i a_j.$ The reader should be aware that this is different from the convention 
employed by 
\cite{ConstantinIyer08,ConstantinIyer11}, who take instead $(\grad_x\ba)_{ij}=\partial a_i/\partial x_j$. 

As observed by \cite{kuzmin1983ideal}, the equation 
\be D_t \bm + (\grad_x\bu)\bm=\bzed, \quad \bu=\bP\bm \lb{eq2-3} \ee 
is equivalent to the incompressible Euler equations for $\bu$ when the flow domain is all of Euclidean space. 
We derive a more general result below, so that here we simply observe that in unbounded space  
\be \phi(\bx,t)= - \frac{1}{4\pi} \int d^3x' \ \frac{1}{|\bx-\bx'|} \grad_{x'}\cdot\bm(\bx',t) \lb{eq2-4} \ee 
assuming good decay of $\bm(\bx',t)$ for $|\bx'|\to\infty,$  and thus 
\be  \bu(\bx,t) = \bm(\bx,t) - \frac{1}{4\pi} \int d^3x' \ \frac{\bx-\bx'}{|\bx-\bx'|^3} \grad_{x'}\cdot\bm(\bx',t).  \lb{eq2-5} \ee 
Integration by parts requires care, because of the divergence at $\bx'=\bx$. Removing a small ball 
$|\bx'-\bx|<\epsilon$ in the $\bx'$-integral and taking the limit $\epsilon\to 0$ gives 
\be  \bu(\bx,t) = \frac{2}{3}\bm(\bx,t) + \dashint d^3x' \ {\bf G}(\bx-\bx') \bm(\bx',t), 
\quad G_{ij}({\bf r})=\frac{1}{4\pi}\left(\frac{3r_ir_j}{r^5}-\frac{\delta_{ij}}{r^3}\right).
\lb{eq2-6}  \ee
where the second term is a principal-value integral with respect to the singular kernel. 
As is well-known (\cite{batchelor2000introduction}, section 7.2), 
$ \bu_0(\bx)=\frac{2}{3}{\bf P}\,\delta^3(\bx-\bx_0) + {\bf G}(\bx-\bx_0){\bf P}$ 
is the velocity field of a vortex ring with impulse ${\bf P}=(1/2) \int d^3x\ \bx\btimes \bom(\bx,t)$
which is centered at $\bx_0$ and whose radius is taken to vanish with ${\bf P}$ fixed. The first delta-function term 
in the velocity is necessary to enforce incompressibility but vanishes for $\bx\neq \bx_0.$ Both 
\cite{kuzmin1983ideal} and \cite{oseledets1989new} therefore interpreted $\bm(\bx,t)$ physically 
as the ``vortex-momentum density'' resulting from a distribution $\sum_{n=1}^N {\bf P}_n(t)\delta^3(\bx-\bX_n(t))$
of infinitesimal vortex-rings in the limit as $N\to\infty$. For finite $N,$  variables $\bX_n(t),$ ${\bf P}_n(t),$
$n=1,...,N$ obey Hamiltonian equations of motion derived by \cite{roberts1972hamiltonian}. 

Generalizing the result of \cite{kuzmin1983ideal}, we now show that the equivalence 
of the Kuz'min-Oseledets equation \eqref{eq2-3} with Euler extends to 
wall-bounded flows, where boundary conditions $\bn \cdot \bm=\partial\phi/\partial n$ on $\bm$ imply
the standard conditions of no-flow through the boundary on velocity $\bu$. Indeed, direct substitution of 
$\bm=\bu+\grad_x\phi$ into \eqref{eq2-3} gives
\be D_t\bu + \grad_x\left(D_t\phi+\frac{1}{2}|\bu|^2\right) = \bzed. \lb{eq2-7} \ee
Defining pressure $p$ up to a space-independent constant $c(t)$ by 
\be D_t\phi + \frac{1}{2}|\bu|^2 -p =c(t) \lb{eq2-8} \ee
then gives 
\be D_t\bu = -\grad_x p \lb{eq2-9} \ee
which has the form of the incompressible Euler equations with kinematic pressure $p.$
The constraint of incompressibility $\grad_x\cdot \bu=0$ implies the standard Poisson equation 
\be -\triangle_x p = {\rm Tr}((\grad_x\bu)^\top(\grad_x\bu)). \lb{eq2-10} \ee
Furthermore, applying $\partial/\partial n$ to the Eq.(\ref{eq2-8}) defining $p$ and using $\partial\phi/\partial n=\bn\cdot\bm$
gives 
\be \left.\frac{\partial p}{\partial n}\right|_{\partial\Omega} =\grad_x\bn: \bu\bu + \bn\cdot[ D_t\bm+ (\grad_x\bu)\bm]. 
\lb{eq2-11} \ee
Together with Eq.\eqref{eq2-3} for $\bm,$ this yields 
\be  \left.\frac{\partial p}{\partial n} \right|_{\partial\Omega} =\grad_x\bn: \bu\bu, \lb{eq2-12} \ee
which are the standard pressure b.c. for incompressible Euler arising from the condition of 
no-flow through the boundary. In order to impose initial conditions $\bu(t=t_0)=\bu_0$
for a suitable choice of $\bu_0,$ satisfying $\grad_x\cdot \bu_0=0$ and $\left.\bn\cdot \bu_0
\right|_{\partial\Omega}=0,$ one may make an arbitrary choice of $\phi_0$ and take 
$\bm_0=\bu_0+\grad_x\phi_0$ as initial data for \eqref{eq2-3}. In that case, the equation 
\eqref{eq2-8} for $\phi$ must be solved with the initial condition 
\be  \phi(t=t_0)=\phi_0. \lb{eq2-13} \ee
Remarkably, initial and boundary conditions for the first-order, hyperbolic PDE \eqref{eq2-8} 
are unconstrained except by the requirement 
of a solution smooth up to the boundary.  Physically, one may interpret different choices of $\phi$ that solve 
this equation to correspond to different sets of image vortices outside $\Omega$ that are needed to produce 
a net incompressible velocity $\bu$ with zero flow though the boundary, for  different momentum 
distributions $\bm$ of vortex-rings inside $\Omega$. In this way the initial-boundary value problem of the Kuz'min-Oseledets 
equation is equivalent to the initial-boundary-value problem  of incompressible Euler, except that there 
are infinitely many distinct choices of  $\bm,$ $\phi$ that correspond to the same $\bu,$ $p.$

Using the vector calculus identity $\bu\btimes(\grad_x\btimes\bm)=(\grad_x\bm)\bu-(\bu\cdot\grad_x)\bm,$ 
the Kuz'min-Oseledets equation can be rewritten as 
\be \partial_t\bm = \bu\btimes(\grad_x\btimes\bm) - \grad_x(\bm\cdot\bu). \lb{eq2-14} \ee
Because $\bm$ and $\bu$ differ only by a gradient, they have same curl:
\be  \grad_x\btimes\bm = \grad_x\btimes\bu:=\bom. \lb{eq2-15} \ee
Taking the curl of Eq.\eqref{eq2-3} rewritten as \eqref{eq2-14} then obviously yields 
the standard inviscid Helmholtz equation for the vorticity:
\be \partial_t \bom = \grad_x\btimes(\bu\btimes\bom). \lb{eq2-16} \ee
Yet another form of Eq.\eqref{eq2-3} follows by regarding ${\rm p}_i$ as the components 
of a differential 1-form ${\rm p}_i dx^i,$ in which case the equation becomes 
\citep{oseledets1989new} 
\be \partial_t\bm + \cL_\bu\bm = \bzed, \lb{eq2-17} \ee
for the {\it Lie-derivative} on 1-forms defined by 
\be  \cL_\bu\bm  := (\bu\cdot\grad_x)\bm+ (\grad_x\bu)\bm. \lb{eq2-18} \ee
See \cite{tur1993invariants,besse2017geometric} for introductions to this concept from differential geometry
generalizing the standard material derivative. The form \eqref{eq2-17} of the Kuz'min-Oseledets equation 
makes manifest its remarkable geometric and Lagrangian properties. In fact, as a Lie-transported 
1-form, vortex-momentum $\bm$ can be represented in terms of initial data its $\bm_0$ by 
the Lagrangian ``back-to-labels" map $\bA(\bx,t)$ as    
\be \bm(\bx,t) =(\grad_x\bA)\cdot \bm_0(\bA(\bx,t)) \lb{eq2-19} \ee
As usual, the map $\bA(\bx,t)$ is defined to have zero material-derivative $D_t\bA=\bzed$ 
and satisfies $\bA(\bx,t_0)=\bx$ at the labelling time $t_0.$ Result \eqref{eq2-19} can be 
interpreted as the ``push-forward'' of the differential 1-form under 
the Lagrangian flow \citep{besse2017geometric} and can be verified 
in a pedestrian manner using the matrix differential equation
\be D_t(\grad_x\bA) + (\grad_x\bu)(\grad_x\bA)={\bf O} \lb{eq2-20} \ee
which follows by taking the gradient of $D_t\bA=\bzed.$ By taking a curl of \eqref{eq2-19}, 
one immediately obtains the famous {\it Cauchy formula} (Cauchy, 1815) for the vorticity 
\be \bom(\bx,t) =  (\grad_x\bA)^{\top\,-1} \bom_0(\bA(\bx,t)) 
= \left.(\bom_0(\ba)\cdot \grad_a)\bX\right|_{\bX(\ba,t)=\bx}, \lb{eq2-21} \ee
where $\bX(\ba,t)$ is the standard Lagrangian flow map that satisfies
\be (d/dt)\bX(\ba,t)=\bu(\bX(\ba,t),t), \quad \bX(\ba,t_0)=\ba \lb{eq2-22} \ee 
and which is the inverse map to $\bA(\bx,t).$  To obtain the first equality in \eqref{eq2-21} 
we used the simple identity $\epsilon_{ijk} \frac{\partial A_l}{\partial x_i}  \frac{\partial A_m}{\partial x_j} 
\frac{\partial A_n}{\partial x_k}=\epsilon_{lmn}$ for an incompressible flow.  The Cauchy formula 
can be regarded also as a ``push-forward'' of the vorticity as a differential 2-form \citep{besse2017geometric}.
This result succinctly expresses the ``frozen-in'' property of vorticity for smooth Euler solutions. 
In particular,  the inverse formula to \eqref{eq2-21} 
\be  \bom_0(\ba) =  (\grad_a\bX)^{\top\,-1} \bom(\bX(\ba,t),t) 
= 
\left.(\bom(\bx,t)\cdot\grad_x)\bA(\bx,t)\right|_{\bA(\bx,t)=\ba} \lb{eq2-23} \ee
defines the {\it Cauchy invariants}, which are independent of time $t$ for every choice of particle label $\ba,$
thus comprising an infinite set of Lagrangian conserved quantities for Euler.  It is intriguing to note, incidentally, 
that the hyperbolic equation \eqref{eq2-8} for the velocity potential $\phi$ can also be solved in terms 
of Lagrangian flows (method of characteristics) and when $c(t)\equiv 0$ one finds 
\be \phi(\bX(\ba,t),t)=\phi(\ba,t_0) - \int_{t_0}^t ds \left[ \frac{1}{2}|\dot{\bX}(\ba,s)|^2-p(\bX(\ba,s),s)\right]. 
\lb{phi-act} \ee 
Hence, this potential is directly related to the (negative of) the {\it action}, which is locally minimized by 
particle trajectories for incompressible Euler solutions \citep{brenier2003topics}.

It was noted in passing by \cite{kuzmin1983ideal} and developed in more detail by \cite{oseledets1989new} that 
a similar equivalence holds between the equation 
\be D_t \bm + (\grad_x\bu)\bm=\nu\triangle_x\bm , \quad \bu=\bP\bm \lb{eq2-24} \ee
and the incompressible Navier-Stokes equation for $\bu,$ at least for flows in unbounded, Euclidean space.  
This equivalence extends to wall-bounded flows in any spatial domain $\Omega,$ with stick b.c. 
on velocity $\bu$ corresponding to the condition $\bm=\grad_x\phi$ at $\partial\Omega.$ More generally,
one may consider walls moving with tangential velocity $\bu_W$ so that $\bn\cdot\bu_W=0,$ where 
the Leray-Hodge decomposition remains defined as in Eqs.\eqref{eq2-1},\eqref{eq2-2}. 
The stick b.c. $\bu=\bu_W$ then correspond to the condition $\bm=\bu_W+\grad_x\phi$ at $\partial\Omega.$ 
Indeed, as for Euler, direct substitution of $\bm=\bu+\grad_x\phi$ into \eqref{eq2-24} gives
\be  D_t\bu + \grad_x\left(D_t\phi+\frac{1}{2}|\bu|^2-\nu\triangle_x\phi\right) = \nu\triangle_x\bu.  \lb{eq2-25} \ee
Defining pressure $p$ up to a space-independent constant $c(t)$ by 
\be D_t\phi -\nu\triangle_x\phi+ \frac{1}{2}|\bu|^2 -p =c(t) \lb{eq2-26} \ee
then gives 
\be D_t\bu = -\grad_x p +\nu\triangle_x\bu \lb{eq2-27} \ee
which has the form of the incompressible Navier-Stokes equations. The constraint of incompressibility
implies the standard Poisson equation for the pressure $p$, just as for Euler. Similarly, 
applying $\partial/\partial n$ to the Eq.\eqref{eq2-26} defining $p$ and using $\partial\phi/\partial n=\bn\cdot\bm,$
\be  \left.\frac{\partial p}{\partial n} \right|_{\partial\Omega} = \grad_x\bn: \bu_W\bu_W +\bn\cdot\nu\triangle_x \bu
+ \bn\cdot[ D_t\bm -\nu\triangle_x\bm + (\grad_x\bu)\bm ].  \lb{eq2-28} \ee
Together with the viscous Kuz'min-Oseledets equation \eqref{eq2-24} for $\bm,$ this yields 
\be \left. \frac{\partial p}{\partial n}\right|_{\partial\Omega}  =
\grad_x\bn: \bu_W\bu_W + \bn\cdot\nu\triangle_x \bu,  \lb{eq2-29} \ee
which are the standard pressure b.c. for incompressible Navier-Stokes. 
Just as for Euler, any solenoidal initial condition $\bu(t=t_0)=\bu_0$ for Navier-Stokes can be obtained by solving 
the Kuzmin-Oseledets equation \eqref{eq2-24} with the corresponding initial condition $\bm(t=t_0)=\bm_0:=\bu_0+\grad_x\phi_0$ for
{\it any} choice of $\phi_0.$ In that case, the parabolic linear equation \eqref{eq2-26} must be 
solved with initial condition $\phi(t=t_0)=\phi_0$ but with any choice of boundary conditions $\phi=\phi_W$ 
on $\partial \Omega$ (and any choice of the constant $c(t)$). In that case,  $\bm=\bu+\grad_x\phi$
will give the corresponding solution of \eqref{eq2-24} with the boundary condition  $\bm=\bu_W+\grad_x\phi$
on $\partial\Omega.$ Just as for the inviscid case, the initial-boundary value problem of the viscous
Kuz'min-Oseledets equation \eqref{eq2-24} is equivalent to the initial-boundary-value problem 
of the incompressible Navier-Stokes equation, but with infinitely many distinct choices of $\bm,$ $\phi$ corresponding 
to the same $\bu,$ $p.$


Using the vector calculus identity $\grad_x\btimes(\grad_x\btimes\bm)=\grad_x(\grad_x\cdot\bm)
-\triangle\bm,$ the viscous Kuz'min-Oseledets equation \eqref{eq2-24} can be rewritten in the same 
manner as the inviscid equation: 
\be \partial_t\bm + \grad_x(\bm\cdot\bu-\nu\grad_x\cdot\bm)= \bu\btimes(\grad_x\btimes\bm) 
- \nu\grad_x\btimes(\grad_x\btimes\bm) . \lb{eq2-30} \ee
Since it remains true that $\bm$ and $\bu$ have the same curl, taking the curl of \eqref{eq2-30} 
immediately yields the viscous Helmholtz equation for the vorticity   
\be \partial_t \bom = \grad_x\btimes(\bu\btimes\bom)+ \nu \triangle_x \bom, \lb{eq2-31} \ee 
which, of course, follows also from the Navier-Stokes equation. Unlike for Euler equations, however, 
it has not been so obvious when viscosity is non-vanishing how to exploit the Kuz'min-Oseledets 
formulation in order to obtain simple geometric and Lagrangian properties of vorticity. 
\cite{ConstantinIyer08,ConstantinIyer11} have achieved this by exploiting a mathematical representation 
of viscosity in the Lagrangian framework via stochastic Brownian motion, as we now briefly explain. 

\subsection{Constantin-Iyer Stochastic Lagrangian Formulation}\lb{sec:CI} 

As has been noted many places in the mathematics (\cite{freidlin1985functional}, \cite{oksendal2013stochastic}), 
physics (\cite{shraiman1994lagrangian}, \cite{falkovich2001particles}) and engineering 
(\cite{sawford2001turbulent}, \cite{keanini2006random}) literatures, 
Laplacian diffusion can be represented in a Lagrangian formulation by a suitable Brownian motion.
For diffusion of momentum by kinematic viscosity $\nu$, this involves stochastic Lagrangian 
trajectories that, backward in time, may be regarded as generalizations of the deterministic 
``back-to-labels" map. Precisely, $\tbA^s_t(\bx)$ for $\bx\in\Omega$ is defined to satisfy  
\be \hd \tbA^s_t(\bx) = \bu(\tbA^s_t(\bx),s) \, \rmd s + \sqrt{2\nu}\, \hd\bW(s), \quad  s<t; \qquad \tbA^t_t(\bx)=\bx 
\lb{eq2-32} \ee
Here the notation ``$\hd$'' with a hat denotes the backward It$\oo$ differential, which is exactly the time-reverse 
of the standard (forward) It$\oo$ differential, and $\bW(s)$ is a vector Brownian motion/stochastic Wiener process. 
Setting $\nu=0,$ one recovers a deterministic ODE, whose solution $\bA^s_t(\bx)$ is the usual 
``back-to-label'' map, for labelling time $s.$ Thus, $\bA^0_t(\bx)=\bA(\bx,t)$ in the notations 
of the previous section. Note, however, that the random process $\tbA^s_t(\bx)$ is not guaranteed to attain time $s=0$ 
before exiting from the domain. For any point $\bx\in\Omega,$ the stochastic particle trajectory $\tbA^s_t(\bx)$ 
released at time $t$ first hits the boundary $\partial\Omega$ at some finite, random time $s=\tsigma_t(\bx)<t,$ 
evolving backward. The stochastic Lagrangian representations obtained by \cite{ConstantinIyer11} for wall-bounded 
flows involve the ``stopped process'' $\tbA^{s\prime}_t(\bx):=\tbA^{s\vee\tsigma_t(\bx)}_t(\bx),$ which 
halts the stochastic Lagrangian particle as soon as it first hits the boundary, with the mathematical notation 
$s\vee\tsigma_t(\bx):=\max\{s,\tsigma_t(\bx)\}.$ Since the realizations of $\tsigma_t(\bx)$ are almost surely 
non-differentiable in $\bx,$ we always take 
$\grad_x\tbA^{s\prime}_t(\bx):=\left.\grad_x\tbA^{\sigma}_t(\bx)\right|_{\sigma=s\vee\tsigma_t(\bx)}$ below. 

With these notations, \cite{ConstantinIyer11} in their Lemma 5.1 obtained the following 
stochastic Lagrangian representation of vortex momentum $\bm$ solving the viscous 
Kuz'min-Oseledets equation, with $\bE$ denoting average 
over the random Brownian motions: 
\be \bm(\bx,t) =\bE\left[  \grad_x\tbA^{0\prime}_t(\bx) 
\cdot \bm_{dat}\left(\tbA^{0\prime}_t(\bx),0\vee\tsigma_t(\bx)\right)\right].
\lb{eq2-33} \ee 
Here $\bm_{dat}$ represents the initial-boundary data, with $\bm_{dat}(\bx,t)=\bm_0(\bx)$ for $t=0$
and $\bm_{dat}(\bx,t)=\bm_W(\bx,t)$ for $\bx\in\partial\Omega.$ Using the results of the previous subsection,
the stochastic representation of the Navier-Stokes velocity stated in Theorem 3.1 of \cite{ConstantinIyer11} 
follows as a simple corollary: 
\begin{eqnarray}
&& \bu(\bx,t) = \bP\bE\left[  \grad_x\tbA^{0\prime}_t(\bx) 
\cdot \bm_{dat} \left(\tbA^{0\prime}_t(\bx),0\vee\tsigma_t(\bx)\right)\right] \cr
&&\cr
&& \hspace{10pt} = \bE\left[  \grad_x\tbA^{0\prime}_t(\bx) 
\cdot \bm_{dat} \left(\tbA^{0\prime}_t(\bx),0\vee\tsigma_t(\bx)\right)\right] -\grad\phi(\bx,t) 
\lb{eq2-34} \end{eqnarray} 
with $\bm_{0}=\bu_0+\grad_x\phi_0$ at $t=0$ for any choice of $\phi_0$ and 
with $\bm_{W}=\bu_W+\grad\phi$ at $\partial\Omega,$ where $\phi$ solves 
equation \eqref{eq2-26} for initial data $\phi_0$ and an arbitrary choice of b.c. $\phi_W$ on $\partial\Omega.$ 
\cite{ConstantinIyer11} used as initial data $\bm_0=\bu_0$ only and also gave only an existence argument 
for suitable boundary data $\bm_{W},$ but our use of the Kuz'min-Oseledets formalism provides a complete 
and explicit representation for the possible data. An interesting point is that this representation 
implies directly that $\left.\bn\cdot\bom\right|_{\partial\Omega}=0$ for motionless walls, since on the boundary
$\bn\cdot\bom=(\bn\btimes\grad_x)\cdot\bm_{W}=(\bn\btimes\grad_x)\cdot \grad_x\phi=0.$
\cite{ConstantinIyer11} provided also in their Proposition 6.1 a stochastic Lagrangian representation 
of Navier-Stokes vorticity: 
\be
\bom(\bx,t) =\bE\left[  (\grad_x\tbA^{0\prime}_t(\bx))^{\top\,-1} 
\cdot \bom_{dat}\left(\tbA^{0\prime}_t(\bx),0\vee\tsigma_t(\bx)\right)\right],
\lb{eq2-35} \ee
with likewise $\bom_{dat}(\bx,t)=\bom_0(\bx)$ for $t=0$ and $\bom_{dat}(\bx,t)=\bom_W(\bx,t)$ 
for $\bx\in\partial\Omega.$ The boundary data $\bom_W$ for the Navier-Stokes vorticity 
is not provided directly, of course, but only indirectly and non-locally by the stick b.c. 
$\bu=\bu_W$ on the velocity field.  In both of the formulas \eqref{eq2-33} and \eqref{eq2-35}, 
the solutions are represented by averages involving both the initial and boundary data. 
However, in a flow entirely bounded by walls and asymptotically for very large $t$ the event 
$\tsigma_t(\bx)>0$ holds with overwhelming probability, so that the solutions then depend almost 
entirely on the boundary data and almost not at all on the initial data. 

We refer to the paper of \cite{ConstantinIyer11} for detailed proofs of Lemma 5.1 and Proposition 6.1.
Here we give only very briefly the main idea of the argument. To prove Lemma 5.1, one 
obtains an evolution equation in $s$ for $\grad_x\tbA^{s}_t(\bx) \cdot \bm\left(\tbA^s_t(\bx),s\right)$
with $\bm$ a given solution of the viscous Kuz'min-Oseledets equation \eqref{eq2-24}, 
starting at $s=t$ and integrating backward to $s=0\vee\tsigma_t(\bx).$ By taking the 
gradient of the stochastic ODE \eqref{eq2-32} for particle trajectories, one obtains  
\be
\rmd (\grad_x\tbA^s_t) = \left. (\grad_x\tbA^s_t) (\grad_x\bu) \right|_{(\tbA^s_t,s)}\rmd s , \quad  s<t; \qquad \grad_x\tbA^t_t={\bf I}
\lb{eq2-36}  \ee 
which is now a deterministic ODE (along a stochastic trajectory) because the gradient of the Brownian
motion vanishes. Using this result together with the backward It$\oo$ lemma, 
\begin{eqnarray} \lb{eq2-37} 
 \hd \left[\grad_x\tbA^{s}_t \cdot \bm\left(\tbA^s_t,s\right)\right] 
&= & \grad_x\tbA^{s}_t  \left. \left[D_s\bm+(\grad_x\bu)\bm-\nu\triangle_x\bm\right]\right|_{(\tbA^s_t,s)}\rmd s \cr 
&& \hspace{30pt} + \sqrt{2\nu}\, \left.(\grad_x\tbA^{s}_t)\, (\hd\bW(s)\cdot\grad)\bm\right|_{(\tbA^s_t,s)} \\ \cr \nonumber
&=& \sqrt{2\nu}\, (\grad_x\tbA^{s}_t)\, \left.(\hd\bW(s)\cdot\grad)\bm\right|_{(\tbA^s_t,s)} \qquad t>s>0\vee\tsigma_t(\bx)
\end{eqnarray} 
This yields the representation of Lemma 5.1, because the final surviving term on the righthand side is a backward 
martingale with zero average. In fact, we obtain the even stronger result that $\grad_x\tbA^{s}_t(\bx) \cdot \bm\left(\tbA^s_t(\bx),s\right)$ 
is a backward martingale, which is thus stochastically conserved and whose (deterministic) value $\bm(\bx,t)$ 
at $s=t$ is equal to its average value at $s=0\vee\tsigma_t(\bx).$ In the same manner one can readily show for
any solution $\bom$ of the viscous Helmholtz equation \eqref{eq2-31} that one has a backward martingale 
\begin{eqnarray} 
&&  \hd \left[(\grad_x\tbA^{s}_t)^{\top\,-1} \cdot \bom\left(\tbA^s_t,s\right)\right] 
=  (\grad_x\tbA^{s}_t)^{\top\, -1}   \left. \left[D_s\bom+(\bom\cdot\grad_x)\bu-\nu\triangle_x\bom\right]\right|_{(\tbA^s_t,s)}\rmd s \cr 
&& \hspace{130pt} + \sqrt{2\nu}\, (\grad_x\tbA^{s}_t)^{\top\,-1} \, \left.(\hd\bW(s)\cdot\grad)\bom\right|_{(\tbA^s_t,s)} \\ \cr \nonumber 
&& \hspace{100pt} =\sqrt{2\nu}\, (\grad_x\tbA^{s}_t)^{\top\, -1} \, \left.(\hd\bW(s)\cdot\grad)\bom\right|_{(\tbA^s_t,s)}  \qquad t>s>0\vee\tsigma_t(\bx)
\lb{eq2-38} \end{eqnarray} 
and this yields the representation for vorticity in Proposition 6.1.
 
\begin{figure}
  \centering
  \includegraphics[width=0.9\textwidth]{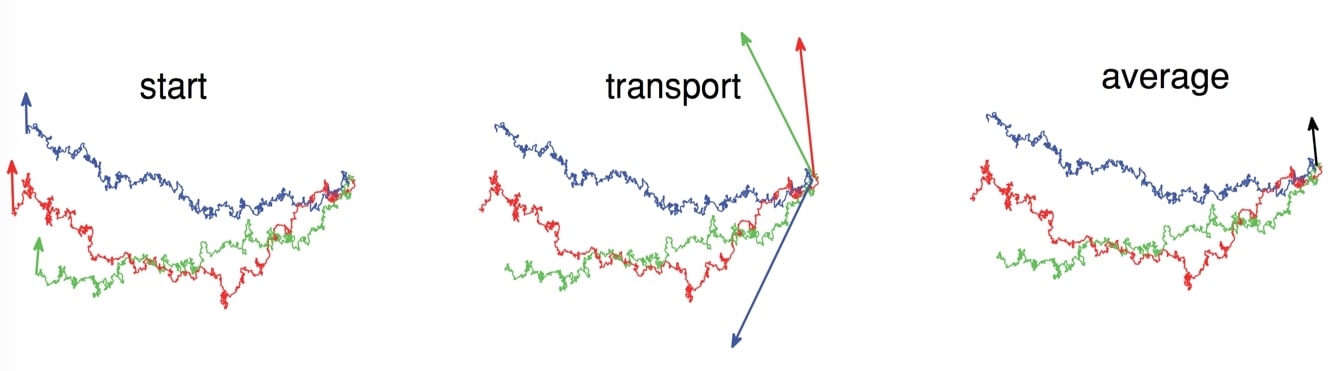}
  \caption{Illustration of the stochastic representation, with three sample realizations shown in
  \textcolor{RoyalBlue}{blue},  \textcolor{red}{red},  and \textcolor{ForestGreen}{green}. 
  {\it Left:} Input vectors at time $\tau=0\vee \tsigma_t(\bx)$ are sampled at random locations by stochastic
  Lagrangian trajectories going backward in time $\tau$ from space-time point $(\bx,t)$. {\it Middle:} The vectors are Lie-transported
  along the stochastic trajectories forward in time from $\tau=0\vee \tsigma_t(\bx)$ to $\tau=t,$ when the vectors arrive 
  to point $\bx$ stretched and rotated by the flow. {\it Right:}  The random vectors at $(\bx,t)$ are ensemble-averaged to obtain
  the resultant vector.}  \label{fig:CauchyFigure}
\end{figure}

The intuitive meaning of these formulas becomes more clear if one integrates along 
stochastic trajectories $\tbA^\tau_t(\bx)$ forward in time from $\tau=0\vee\tsigma_t(\bx)$ to $\tau=t,$ rather  
than backward in time. Note that an explicit solution for $\grad_x\tbA^s_t$ is given 
by the time-ordered matrix exponential, with matrix products ordered right to left for increasing time: 
\be \grad_x\tbA^s_t = {\rm Texp}\,\left[-\int_s^t dr\ \left.(\grad_x\bu)\right|_{(\tbA^r_t,r)}\right].  \lb{eq2-39}  \ee
If one defines starting values in terms of the randomly sampled initial-boundary data  
\be \tilde{\bm}(\bx,0\vee\tsigma_t(\bx)) := \bm_{dat}\left(\tbA^{0\prime}_t(\bx),0\vee\tsigma_t(\bx)\right), 
\lb{eq2-40} \ee 
and defines at later times 
\be \tilde{\bm}(\bx,\tau) := {\rm Texp}\,\left[-\int_{0\vee\tsigma_t(\bx)}^\tau dr\ \left.(\grad_x\bu)\right|_{(\tbA^r_t,r)}\right] 
\tilde{\bm}(\bx,0\vee\tsigma_t(\bx)), \quad \quad t>\tau > 0\vee\tsigma_t(\bx), \lb{eq2-41} \ee 
then the stochastic representation in Lemma 5.1 can be expressed simply as 
\be \bm(\bx,t) = \bE\left[\tilde{\bm}(\bx,t)\right]. \lb{eq2-42} \ee 
However, by direct differentiation of the matrix exponential in \eqref{eq2-41} one can see that 
\be \frac{d}{d\tau} \tilde{\bm}(\bx,\tau) = -\left. (\grad_x\bu)\right|_{(\tbA^\tau_t,\tau)} \tilde{\bm}(\bx,\tau), \quad\quad 
 t>\tau > 0\vee\tsigma_t(\bx). \lb{eq2-43} \ee
This appears exactly the same as the Lie-transport equation of a 1-form except that the material 
derivative $D_\tau$ has been replaced with the time-derivative $\partial/\partial \tau$ along the 
random trajectory $\tbA^\tau_t(\bx)$ starting at $\tau= 0\vee\tsigma_t(\bx)$ and ending at $\tau=t.$
We thus see that the randomly sampled initial-boundary data $\tilde{\bm}(0\vee\tsigma_t(\bx))$
is quasi-Lie-transported as a 1-form along stochastic trajectories forward in time to point $(\bx,t)$
and the resulting contributions $\tilde{\bm}(\bx,t)$ are averaged over the ensemble to yield $\bm(\bx,t).$
The process is illustrated in Figure \ref{fig:CauchyFigure}. 
In this stochastic Lagrangian representation, the quasi-Lie-transport along stochastic trajectories represents 
the nonlinear dynamics of vortex-momentum under the Kuz'min-Oseledets equation and cancellations 
in the ensemble average represent the destruction of vortex-momentum by viscosity. The analogous 
interpretation holds for the vorticity representation in Proposition 6.1. Using the similar matrix exponential 
\be (\grad_x\tbA^s_t)^{\top\,-1}  = {\rm Texp}\,\left[ \int_s^t dr\ \left.(\grad_x\bu)^\top\right|_{(\tbA^r_t,r)}\right],
\lb{eq2-44}  \ee
one can show that randomly sampled initial-boundary data  
\be 
\tilde{\bom}(\bx,0\vee\tsigma_t(\bx) ) = \bom_{dat}\left(\tbA^{0\prime}_t(\bx),0\vee\tsigma_t(\bx)\right)
\lb{eq2-45} \ee
are quasi-Lie-transported along stochastic trajectories $\tbA^\tau_t(\bx)$ from $\tau= 0\vee\tsigma_t(\bx)$ to $\tau=t$
according to the equation 
\be 
\frac{d}{d\tau} \tilde{\bom}(\bx,\tau) = \left. (\tilde{\bom}(\bx,\tau)\cdot\grad_x)\bu \right|_{(\tbA^\tau_t,\tau)}, \quad
\quad  t>\tau > 0\vee\tsigma_t(\bx), 
\lb{eq2-46} \ee 
which closely resembles the usual inviscid evolution equation for vorticity. The final contributions at $\tau=t$ are then 
averaged over the ensemble to give the resultant vorticity
\be 
\bom(\bx,t) = \bE\left[\tilde{\bom}(\bx,t)\right], 
\lb{eq2-47} \ee
and cancellations in this average over random contributions represent  the viscous destruction 
of vorticity in a Lagrangian framework.   

Although the physical interpretation is made more transparent by integrating forward in time, 
the method of proof integrating backward in time yields a more powerful result. It is obviously 
an arbitrary decision to make $0$ the ``initial time'' and one can instead solve the viscous 
Kuz'min-Oseledets and Helmholtz equations starting at any time $s<t$ with initial data $\bm(\cdot,s).$ 
In that case, the same arguments as for $s=0$ yield  
\be 
\bm(\bx,t) = \bE\left[ \tilde{\bm}_s(\bx,t)\right] , \quad s<t  \lb{eq2-48} \ee
with 
\be 
\tilde{\bm}_s(\bx,t) := \grad_x\tbA^{s\prime}_t(\bx) 
\cdot \bm\left(\tbA^{s\prime}_t(\bx),s\vee\tsigma_t(\bx)\right), 
\lb{eq2-49} \ee
and likewise 
\be 
\bom(\bx,t) =\bE\left[ \tilde{\bom}_s(\bx,t)\right], \quad s<t \lb{eq2-50} \ee
with
\be 
\tilde{\bom}_s(\bx,t):=(\grad_x\tbA^{s\prime}_t(\bx))^{\top\,-1} 
\cdot \bom\left(\tbA^{s\prime}_t(\bx),s\vee\tsigma_t(\bx)\right). 
\lb{eq2-51} \ee
The quantities $\tilde{\bm}_s(\bx,t)$ and $\tilde{\bom}_s(\bx,t)$ defined above 
are the backward martingale random processes in $s$ that were used in the proofs
of Lemma 5.1 and Proposition 6.1, which are thus statistically conserved for every choice of $\bx\in\Omega.$
If $s=0,$  then $\tilde{\bm}_0(\bx,t),$ $\tilde{\bom}_0(\bx,t)$
are the same quantities $\tilde{\bm}(\bx,t),$ $\tilde{\bom}(\bx,t)$ that were introduced in discussing 
the physical interpretation using forward integration.
The proofs in \cite{ConstantinIyer11} thus provide an infinite set of stochastic Lagrangian conservation laws  
for incompressible Navier-Stokes solutions. We shall refer to $\tilde{\bom}_s(\bx,t)$ for $\bx\in\Omega$
and $s<t,$ in particular, as the {\it stochastic Cauchy invariants}, since they generalize the Cauchy vorticity 
invariants of incompressible Euler solutions. One fundamental difference is that the Cauchy invariants for Euler 
are conserved both forward and backward in time, but the stochastic Cauchy invariants for Navier-Stokes 
are conserved only backward in time. This difference arises because the Navier-Stokes equation, unlike 
Euler, is time-irreversible and the stochastic Cauchy invariants express the fixed arrow of time. As we shall see
in the discussion of numerical methods in the next section, realizations of $\tilde{\bm}_s(\bx,t)$, 
$\tilde{\bom}_s(\bx,t)$ can be calculated for all $s<t$ by a simple integration scheme backward in time
using ensembles of stochastic Lagrangian particles. The requirement that their averages values remain 
independent of $s$ is a stringent test on the accuracy of the numerics. 


\subsection{Relation with the Lighthill-Morton Theory}\lb{sec:LM}  

The \cite{ConstantinIyer11} representation \eqref{eq2-50}, \eqref{eq2-51} for any flow entirely bounded by walls 
allows one to reconstruct the interior vorticity $\bom(\bx,t)$
at any point $\bx\in\Omega$ and time $t$ in terms of the vorticity at the boundary, by taking the time 
interval $t-s$ so large that $\tsigma_t(\bx)>s$ with overwhelming probability. 
This formula thus makes very concrete the truism that all vorticity in the flow originates, 
ultimately, at a solid wall. The \cite{ConstantinIyer11} theory does not address the generation of the wall-vorticity, 
but instead takes the boundary values $\bom_W(\bx,t)$ for $\bx\in\partial\Omega$ as a given. It therefore plays 
a rather complementary role to the theory of \cite{lighthill1963boundary} and \cite{morton1984generation},  
which was advanced to explain the inviscid generation of vorticity at the wall.
The \cite{ConstantinIyer11} formulas instead describe, in a Lagrangian framework, the subsequent transport of 
vorticity away from the walls by nonlinear advection, stretching and viscous diffusion. In order to discuss more 
clearly the relations of the two approaches,  we must first review briefly the Lighthill-Morton theory, especially since some 
minor controversies still exist regarding its general formulation.  

The concept of ``vorticity source'' or ``vorticity flux density'' was first introduced in the pioneering work of 
\cite{lighthill1963boundary}, who identified vorticity generation in an infinitesimal layer at a solid, stationary 
wall as an essentially inviscid process driven by tangential pressure gradients. Lighthill's theory was extended 
by \cite{morton1984generation} to accelerating walls, with the conclusion that the source of 
tangential vorticity at the wall is given by 
\be 
\bsigma = \left. \bn\btimes (\grad_xp + D_t\bu) \right|_{\partial\Omega} 
\lb{eq2-52} \ee
at least for a flow domain with a two-dimensional flat boundary and constant outward unit normal vector, say,
$\bn=-\hat{{\bf y}}.$ This quantity has dimensions of (vorticity)$\times$(velocity). Concretely, 
by the Kelvin theorem, $\bsigma\cdot \bn\btimes d{\bf l}$ represents the rate of generation 
of circulation along a small line-element $d{\bf l}=\hat{{\bf t}} ds,$ with $\hat{{\bf t}}$ any unit vector tangent 
to the wall, and thus $\bsigma\cdot \bn\btimes\hat{{\bf t}}$ represents the wall-normal flux of 
vorticity generated in the direction $\bn\btimes\hat{{\bf t}}$ \citep{morton1984generation,eyink2008turbulent}. 
Applying the momentum-balance (Navier-Stokes) equations 
at the wall, for the simplified two-dimensional geometry, \cite{lighthill1963boundary} and \cite{morton1984generation} 
obtained alternative expressions for tangential vorticity source: 
\be
\sigma_x =  \left.-\nu \frac{\partial\omega_x}{\partial y}\right|_{\partial\Omega}, \quad
\sigma_z =  \left.-\nu \frac{\partial\omega_z}{\partial y}\right|_{\partial\Omega}. 
\lb{eq2-53} \ee
These formulas express the instantaneous balance between rate of generation of vorticity at the wall and 
the rate of viscous diffusion of vorticity away from the wall. Using the facts that 
$\left.\partial\omega_y/\partial x\right|_{\partial\Omega}=0$ and 
$\left.\partial\omega_y/\partial z\right|_{\partial\Omega}=0,$  the 
latter expressions can be rewritten as 
\be
  \bsigma  = \left. -\nu \bn\btimes (\grad_x\btimes\bom) \right|_{\partial\Omega}
=  \left. \bSigma\bn\right|_{\partial\Omega}, 
\lb{eq2-54} \ee
where $\bSigma$ is the vorticity flux for Navier-Stokes defined in \eqref{eq2} as an anti-symmetric matrix. 
Since $-\bSigma^\top\bn$ represents generally the flux of vector vorticity in the space direction $-\bn$ (inward unit normal),
the formula \eqref{eq2-54} has a natural physical interpretation and plausibly generalizes the 
Lighthill-Morton theory to flow domains with three-dimensional, curvilinear boundaries. 
This generalization was first proposed by \cite{lyman1990vorticity}. 

There is a subtlety, however, because the concept of ``vorticity flux vector'' is only defined up to a divergence-free contribution,
and the most standard choice of viscous flux is not the anti-symmetric tensor in \eqref{eq2}  but is instead 
$\bSigma'_{visc} =-\nu\grad_x\bom.$ This yields 
\be \bsigma' =  -\bSigma^{\prime \top} \bn \big|_{\partial\Omega} = \nu (\bn\cdot\grad_x)\bom \big|_{\partial\Omega}, 
\lb{eq2-55} \ee
which was suggested already by \cite{panton1984incompressible}, sections 13.7, 13.11, 
and still often espoused \citep{wu1993interactions, wu1996vorticity, wu1998boundary}. 
There are obvious distinctions between the two proposals. In particular, $\bSigma$ is anti-symmetric, so that 
$\bsigma\cdot\bn=0$ and thus Lyman's generalization predicts no generation of vorticity normal to the boundary. 
However, even for a flat, two-dimensional wall, typically $\bsigma'\cdot\bn\neq 0.$
The latter result seems implausible, since wall-normal vorticity is associated to flux through loops lying entirely within 
the wall and pressure-gradients can drive no net circulation round such loops. However, we believe that 
a stronger argument in favor of $\bsigma$ rather than $\bsigma'$, is that it is $\bSigma$ which 
appears in the momentum balance \eqref{eq1} and not $\bSigma'.$ Thus, it is only \eqref{eq2-54} which corresponds to 
the Lighthill-Morton inviscid expression \eqref{eq2-52} for a general three-dimensional boundary, and the Kelvin-theorem argument of 
\cite{morton1984generation} and \cite{eyink2008turbulent} then gives a spatially local meaning to $\bsigma.$ 
For this reason, we shall here adopt \eqref{eq2-52} and \eqref{eq2-54} as the generally correct formulation.   
Fortunately there is no difference in the predictions of the competing proposals of \cite{lyman1990vorticity} and 
\cite{panton1984incompressible} for the example which is studied numerically in this work, namely, 
the flux of tangential vorticity away from a flat, two-dimensional channel wall, so we may disregard this issue hereafter. 

The Lighthill-Morton theory is directly related to the statistical result \eqref{eq4} derived by \cite{taylor1932transport}  
and \cite{huggins1994vortex} for statistically stationary turbulence in domains with non-accelerating walls:
\be \overline{\Sigma}_{ij} = \epsilon_{ijk} \partial_k \overline{\left( p+\frac{1}{2}|\bu|^2\right)} 
\lb{eq2-56} \ee
which locally relates vorticity flux and generalized pressure-gradients at interior points.  The time-averaged 
tangential vorticity source relation of Lighthill-Morton:  
\be \overline{\sigma}_i =\overline{\Sigma}_{ij}\hat{n}_j\big|_{\partial\Omega}  
= (\bn\btimes\grad_x \overline{p})_i \big|_{\partial\Omega} \lb{eq2-57} \ee
is just the Taylor-Huggins relation \eqref{eq2-56} approaching a boundary point and considering wall-normal flux of vorticity 
driven by mean tangential pressure-gradients. Here we note that there may be an average transport of tangential vorticity 
also parallel to the wall, driven by mean wall-normal gradients of generalized pressure:
\be \frac{\partial}{\partial n} \overline{p+(1/2)|\bu|^2}\big|_{\partial\Omega}  = \frac{1}{2} \epsilon_{ijk} \hat{n}_i \overline{\Sigma}_{jk}\big|_{\partial\Omega} 
=-\nu(\bn\btimes\grad_x)\cdot\overline{\bom}\big|_{\partial\Omega}. \lb{eq2-58} \ee
The above relation is just the time-average of the standard Neumann boundary condition on the pressure 
(noting that $(\partial/\partial n)|\bu|^2=0$ at the wall).  The Taylor-Huggins result \eqref{eq2-56} continues both of the time-averaged 
relations \eqref{eq2-57} and \eqref{eq2-58} into the interior of the flow, 
where the vorticity current tensor in \eqref{eq2} for an incompressible Navier-Stokes fluid takes the form:  
\be \Sigma_{ij} = u_i\omega_j - u_j\omega_i + \nu\left(\frac{\partial\omega_i}{\partial x_j}-
\frac{\partial\omega_j}{\partial x_i}\right),  \lb{eq2-59} \ee
with contributions not only from viscous diffusion but also from nonlinear advection and stretching by the fluid flow velocity. 

%
%
%
%
%
%
%

The Constantin-Iyer formalism provides a Lagrangian description of the space-transport of vorticity,
with contributions from the same physical processes that are represented by the Eulerian 
vorticity current \eqref{eq2-59}. The stochastic Cauchy invariant \eqref{eq2-51} can be directly 
related to the vorticity current by integration over any open subdomain $O\subset \Omega$ with 
smooth boundary $\partial O$ and by statistical expectation, yielding 
\be \bE\left[\int_O d^3x \Big(\tilde{\bom}_s(\bx,t) - \bom(\bx,s) \Big) \right] =  \int_s^t dr 
\int_{\partial O} \bSigma(\bx,r)\cdot d{\bf A}, \quad s<t, \lb{eq2-60} \ee
where $d\bA$ is the outward-pointing vector area element on $\partial O.$ Of course, here $\bSigma$ 
and $\bSigma'$ yield the same total surface integral, as does any other vorticity current 
that differs from these two by a divergence-free vector. Detailed understanding of the physical processes 
contributing in the Constantin-Iyer representation can be obtained by decomposing 
\be \Big(\tilde{\bom}_s(\bx,t) - \bom(\bx,s) \Big)  = \left(\bom(\tbA^s_t(\bx),s)-\bom(\bx,s)\right) 
+ \left(\tilde{\bom}_s(\bx,t) -\bom(\tbA^s_t(\bx),s)\right). \lb{eq2-61} \ee
The first term on the right of \eqref{eq2-61} 
represents spatial transport of vorticity by fluid advection and viscous diffusion,
while the second term represents space transport by vortex-stretching. To see this, note that the 
space-integral of the first term gives 
\be
\int_O d^3x\ \left(\bom(\tbA^s_t(\bx),s)-\bom(\bx,s)\right) =\int_{\tbA^s_t(O)\backslash O} d^3x\ \bom(\bx,s)
- \int_{O\backslash \tbA^s_t(O)} d^3x\ \bom(\bx,s) \lb{eq2-62} \ee
for $t>s>\tsigma_t(O):=\sup_{\bx\in O} \tsigma_t(\bx),$ because the stochastic flow preserves the volume 
elements in $O$ before any particle hits the flow boundary. The expression on the  right side of \eqref{eq2-62} 
transparently represents change of the space-integral by transport due to advection and viscous
diffusion. Indeed, the integration set $\tbA^s_t(O)\backslash O$ consists of points outside $O$ at time $s$ 
that enter it by time $t$ and their contribution has a positive sign, whereas the set $O\backslash \tbA^s_t(O)$ 
consists of points inside $O$ at time $s$ that leave it by time $t$ and their contribution has a negative sign. 
Note furthermore from \cite{ConstantinIyer11}, Proposition 5.2 that 
\begin{eqnarray} 
\lim_{s\to t-} \frac{1}{t-s} \bE\left[ \bom(\tbA^{s}_t(\bx),s)-\bom(\bx,s)\right] &=& 
    -(\bu(\bx,t)\cdot\grad)\bom(\bx,t) + \nu\triangle\bom(\bx,t)   \cr
   &=& -\grad\cdot \big[\bu\bom +\bSigma_{visc} \big] 
\lb{eq2-63}    
\end{eqnarray} 
with $ \bSigma_{visc} = - \nu\left[(\grad\bom)-(\grad\bom)^\top\right]$. 
The first term in \eqref{eq2-61} thus recovers the instantaneous contributions to $\bSigma$ from 
advection and viscous diffusion in the limit $s\to t-.$ On the other hand, the second term in \eqref{eq2-61} 
can be rewritten using the definition \eqref{eq2-51} of the stochastic Cauchy invariant, as:  
\be \left(\tilde{\bom}_s(\bx,t) -\bom(\tbA^s_t(\bx),s)\right) =
\left[(\grad_x\tbA^{s}_t(\bx))^{\top\,-1} -{\bf I}\right] \cdot \bom\left(\tbA^{s}_t(\bx),s\right), \lb{eq2-64} \ee
for $t>s>\tsigma_t(\bx).$ The vector $\bom\left(\tbA^{s}_t(\bx),s\right)$ represents the initial 
vorticity sampled at time $s,$ while the matrix $(\grad_x\tbA^{s}_t(\bx))^{\top\,-1} -{\bf I}$ represents 
the accumulated change from stretching and rotation between times $s$ and $t,$ as illustrated 
by the middle panel of Figure \ref{fig:CauchyFigure}. Note that it follows either from 
the equation of motion \eqref{eq2-36} for $\grad_x\tbA^s_t(\bx)$ or from the closed expression 
\eqref{eq2-44} in terms of a time-ordered exponential, that 
\begin{eqnarray} 
\lim_{s\to t-} \frac{1}{t-s} 
\left((\grad_x\tbA^{s}_t(\bx))^{\top\,-1} -{\bf I}\right) \cdot \bom\left(\tbA^{s}_t(\bx),s \right)
&= & (\bom(\bx,t)\cdot\grad)\bu(\bx,t), \cr
&=& -\grad\cdot \big[-\bom\bu\big]. 
\lb{eq2-65} 
\end{eqnarray} 
Thus, the second term in \eqref{eq2-61} recovers the instantaneous vortex-stretching contribution 
to $\bSigma$ in the limit $s\to t-$. Combining \eqref{eq2-63} and \eqref{eq2-65} gives 
\be \lim_{s\to t-} \frac{1}{t-s} \bE\big[ \tilde{\bom}_s(\bx,t)-\bom(\bx,s)\big] = -\grad\cdot\bSigma, 
\lb{eq2-66} \ee
which is a spatially local version of \eqref{eq2-60} for infinitesimal times. 

While the analysis of \cite{ConstantinIyer08,ConstantinIyer11} takes the vorticity at the wall as Dirichlet boundary 
data for the Helmholtz equation,  a more direct connection with the Lighthill-Morton theory can be established 
by taking instead the vorticity source density \eqref{eq2-54}-\eqref{eq2-55} as Neumann data. This generalization 
follows by utilizing stochastic Lagrangian trajectories reflected at the wall and the concept of 
``boundary local-time density'' as in \cite{drivas2017lagrangianII}, but the mathematical derivation 
and numerical implementation must be deferred to a following work \citep{eyink2020stochastic}.  
In the application to channel-flow turbulence in paper II, 
the preceding discussion 
implies a constant average transport of spanwise vorticity by the stochastic Lagrangian flow in the 
wall-normal direction, with the vorticity flux driven by pressure-drop downstream in the channel.  
One of the main long-term goals of our program of study is a detailed Lagrangian description 
of the dynamical processes contributing to this systematic vorticity transport.  


\section{Numerical Methods and Their Evaluation}\lb{sec:num}  

The mathematical theory discussed above can be implemented numerically in a straightforward fashion.
We outline below a simple algorithm to calculate realizations of the stochastic Cauchy invariants and to evaluate 
their statistical averages, given the velocity and velocity-gradient fields associated to any solution of the 
incompressible Navier-Stokes equation. This algorithm is then implemented with an online database 
of a turbulent channel-flow solution \citep{graham2016web}, validating our computational method by verifying 
the conservation in the mean of the Cauchy vorticity invariant. In the following paper II we shall then exploit 
the stochastic Cauchy invariant to study numerically the Lagrangian vorticity dynamics for a couple of selected 
events in the buffer-layer of the channel-flow, allowing us to explicate in detail the origin of the vorticity at the wall.

\subsection{Numerical Monte Carlo Lagrangian Scheme}\lb{sec:time-int}  

The numerical method we will elaborate is based on a straightforward time-discretization of the 
stochastic differential equations \eqref{eq2-32} for the ``back-to-labels'' maps, or the histories 
of stochastic Lagrangian particles backward in time.  We use the Euler-Maruyama method to calculate 
stochastic particle positions $\tbA^{s}_t=(\ta^{s}_t, \tb^{s}_t, \tc^{s}_t)$ 
over discrete times $s=s_k:=t-k(\Delta s),$ $k=0,1,2,...$ by backward integration:  
\be \tbA^{s_k}_t(\bx) = \tbA^{s_{k-1}}_t(\bx) - \bu(\tbA^{s_{k-1}}_t(\bx),s_{k-1}) \,  \Delta s + \sqrt{2\nu\Delta s}\, \tilde{{\bf N}}^k, \quad  
k=1,2,3,... ; \quad \tbA^t_t(\bx)=\bx. \lb{eq3-1} \ee 
See \cite{kloeden2013numerical}.
Here $\tilde{{\bf N}}^k$ is a three-dimensional normal random vector with mean zero and covariance matrix ${\bf I},$ independently 
sampled for each step $k=1,2,3,...$. For additive noise,
as in the equation \eqref{eq3-1}, the Euler-Maruyama scheme is first-order strong convergent as $\Delta s\to 0.$ 
We use the Mersenne Twister algorithm \citep{matsumoto1998mersenne,matsumoto1997mersenne}
to produce a pseudo-random sequence of uniform random  numbers and we generate samples of normal random 
numbers pairwise using the transform method of \cite{box1958note}. To numerically approximate the Cauchy invariant 
as defined in \eqref{eq2-51}, or 
\be \tilde{\bom}_s(\bx,t) = (\grad_x\tbA^{s\prime}_t(\bx))^{-1\,\top}\bom(\tbA^{s\prime}_t(\bx),s\vee\tsigma_t(\bx)) \lb{eq3-2} \ee 
we must also discretize the evolution equation for the {\it deformation gradient tensors} $\tbD^s_t(\bx) :=(\grad_x\tbA^s_t)^{-1\,\top},$ 
or the Jacobian derivative matrices of the flow map. The equation for $\tbD^s_t(\bx)$ is easily derived from the equation 
\eqref{eq2-36} for $\grad_x\tbA^s_t(\bx)$ to be
\be \rmd \tbD^s_t = \left. -\tbD^s_t \cdot (\grad_x\bu)^\top \right|_{(\tbA^s_t,s)}\rmd s , 
\quad  s<t; \qquad \tbD^t_t={\bf I} \lb{eq3-3} \ee 
with the corresponding Euler approximation 
\be \tbD^{s_k}_t = \tbD^{s_{k-1}}_t \cdot \left[{\bf I}  + 
\left. (\grad_x\bu)^\top\right|_{(\tbA^{s_{k-1}}_t,s_{k-1})} \Delta s\right],  
\quad  k=1,2,3,...; \qquad \tbD^t_t={\bf I}. \lb{eq3-4} \ee 

To calculate averages, a statistically independent ensemble of $N$ stochastic Lagrangian particles is evolved, 
with positions $\tbA^{s_k,(n)}_{t}(\bx)$ and deformation matrices $\tbD^{s_k,(n)}_t(\bx)$ for $n=1,...,N.$   
At every $k$th-step in time $s$, Navier-Stokes solution fields $\bu$ and $\grad_x\bu$ must be evaluated at the current set 
of particle locations and used to evolve both them and the deformation matrices backward to the next time. 
At each time-step $k$, the algorithm checks whether the $n$th particle has left the domain:
\be  |\tb^{s_k,(n)}_{t}(\bx)|>h.  \lb{eq3-5} \ee 
If so, this $n$th particle is declared to have been ``born'' at that time and evolved no further backward. 
Interpolation is used to identify the wall-hitting time 
$s=\tsigma_{*}^{(n)}=s_{k-1}-\Delta\tsigma_{*}^{(n)}$ at which the condition $|\tb^{s,(n)}_{t}(\bx)|=h$ first occurred.
Note that $\tsigma_{*}^{(n)}$ is a random quantity, even given both $\tbA^{s_{k-1},(n)}_{t}(\bx)$ and $\tbA^{s_k,(n)}_{t}(\bx).$ 
As discussed in Appendix \ref{sec:interp}, this hitting time can be estimated by a stochastic interpolation method 
that becomes exact in the limit of a vanishing wall-normal velocity, with no restriction whatsoever 
on the time-step $\Delta s.$ The same interpolation scheme gives also the wall positions $(a,c)=(\tilde{a}_*^{(n)}, \tilde{c}_*^{(n)})$ 
at the first-hitting time. Furthermore, the deformation matrix is updated to the hitting time by the Euler formula
with a fractional time-step: 
\be  \tbD^{\tsigma_*^{(n)},(n)}_t = {\tbD^{s_{k-1},(n)}_t} \cdot \left[{\bf I}  + 
\left. (\grad_x\bu)^\top\right|_{(\tbA^{s_{k-1},(n)}_t,s_{k-1})} \Delta \tsigma_*^{(n)}\right]   \lb{eq3-6} \ee 
and used to evaluate the Cauchy invariant for the newly ``born'' particle by the formula  
\be \tilde{\bom}_*^{(n)} =   \tbD^{\tsigma_*^{(n)},(n)}_t \cdot \bom(\tilde{a}_*^{(n)}, \tilde{b}_*^{(n)}, 
\tilde{c}_*^{(n)},\tsigma_*^{(n)}), \quad \tilde{b}_*^{(n)}=\,\pm h  \lb{eq3-7} \ee 
requiring the velocity-gradient $\grad_x\bu$ of the Navier-Stokes solution at space-time point
$(\tilde{a}_*^{(n)},\tilde{b}_*^{(n)}, \tilde{c}_*^{(n)},\tsigma_*^{(n)}).$ The random values for each newly 
``born'' particle with label $n,$
\be  \tsigma_*^{(n)}, \ \tilde{a}_*^{(n)}, \ \tilde{c}_*^{(n)}, \ \tilde{\bom}_*^{(n)},  \lb{eq3-8} \ee 
are then output to a file. Subsequently, for all of the remaining particles $n$, 
the so-called ``alive'' ones which have not yet hit the wall, the deformation matrix 
is updated by the Euler formula \eqref{eq3-4} to the new time $s_k=s_{k-1}-\Delta s$ and then that matrix is 
used to calculate an updated Cauchy invariant at time $s_k$ with 
\be \tilde{\bom}_{s_k}^{(n)}(\bx,t)  = \tbD^{s_{k},(n)}_t\cdot \bom(\tbA^{s_k}_t(\bx),s_k), \lb{eq3-9} \ee  
requiring the Navier-Stokes velocity-gradient $\grad_x\bu$ at time $s_k$ to obtain $\bom(\tbA^{s_k}_t(\bx),s_k).$

This entire procedure is repeated in all subsequent time-steps. The set of stochastic 
particles labelled by $n\in {\mathbb N}=\{1,2,...,N\}$ is partitioned into a subset ${\mathbb W}_k$ 
representing ``wall"/``pre-born'' particles that have already hit the wall by time-step $k$ and the complementary 
subset ${\mathbb I}_k={\mathbb N}\backslash {\mathbb W}_k$ representing ``interior'' /``alive'' particles still in the 
flow interior at time-step $k.$ At each step, the variables associated to ``wall'' particles 
with $n\in{\mathbb W}_k$ are left unchanged and only the variables associated to ``interior'' particles 
with $n\in{\mathbb I}_k$ are evolved from time $s_k$ back to time $s_{k+1}.$ Identifying 
the newly ``born'' particles, storing the values of their variables at their first-hitting time, 
and adding their indices $n$ to ${\mathbb W}_k$ then yields the new sets ${\mathbb W}_{k+1}\supseteq {\mathbb W}_k$ 
and ${\mathbb I}_{k+1}={\mathbb N}\backslash {\mathbb W}_{k+1}$ for the next time-step. At every step,  
ensemble-means of random variables $\tilde{F}$ are calculated by averaging over the entire set of particles 
${\mathbb N},$ both wall and interior, as:
\be {\mathbb E}[\tilde{F}] \doteq \frac{1}{N} \sum_{n=1}^N \tilde{F}^{(n)} \lb{eq3-10} \ee 
The convergence rate of this scheme as $N\to\infty$ is rather slow, with standard Monte Carlo errors
by the central limit theorem of order $\sim \sqrt{{\rm Var}[F]/N}$, where ${\rm Var}[F]$
is the variance of the averaged variable. To obtain numbers of samples $N$ sufficiently large, 
we implemented the preceding algorithm in Fortran on a cluster computer at the     
Maryland Advanced Research Computing Center (MARCC). The Monte Carlo averaging 
scheme is perfectly parallelizable, with statistically independent subsets of samples assigned 
to distinct cluster nodes and no communication whatsoever required between nodes.   

An annotated version of the Fortran code which implements the algorithm discussed above is available at the 
Github repository \citep{SCauchy2019}. This code is written to be used with the JHU online channel-flow database, 
which is discussed next.

\subsection{JHU Channel-Flow Database}\lb{sec:JHTDB}

The Johns Hopkins Turbulence Databases (JHTDB) channel-flow dataset \citep{graham2016web} is exploited for the empirical study in this paper.
This data was generated from a Navier-Stokes simulation in a channel using a pseudospectral method in the plane parallel to the 
walls and a seventh-order B-splines collocation method in the wall-normal direction \citep{lee2013petascale}. For numerical solution, the 
Navier-Stokes equations were formulated in wall-normal velocity-vorticity form \citep{kim1987turbulence}. Pressure was computed by 
solving the pressure Poisson equation only when writing to disk, which was every five time steps for 4000 snapshots, enough for about one domain 
flow-through time. The simulation domain $[0,8\pi] \times [-h,h] \times [0,3\pi]$ with channel half-height $h=1$ was discretized using a spatial grid of 
$2048 \times 512 \times 1536$ points in the streamwise ($x$), wall-normal ($y$), and spanwise ($z$) directions, respectively. Time advancement 
was made with a third-order low-storage Runge-Kutta method and dealiasing was performed with 2/3 truncation \citep{orszag1971elimination}. 
A constant pressure head was applied to drive the flow at $Re_\tau = 1000$ ($Re_{bulk} = \frac{2hU_{bulk}}{\nu} = 40\,000$) with 
bulk velocity near unity. As is common, we shall indicate with a superscript ``+'' non-dimensionalized quantities in viscous wall units,
with velocities scaled by friction velocity $u_*$ and lengths by viscous length $\delta_\nu=\nu/u_*=10^{-3}.$ Also as usual, we define 
$y^+=(h\mp y)/\delta_\nu$ near $y=\pm h.$ In these units,
the first $y$-grid point in the simulation is located at distance $\Delta y_1^+ = 1.65199\times 10^{-2}$ from the wall, while in the center 
of the channel $\Delta y_c^+ = 6.15507.$ Other numerical parameters of the channel-flow dataset are summarized in Table 1.

\begin{table}  \label{tab:table1}
    \begin{center}
    \vspace{-10pt} 
    \caption{Simulation Parameters for Turbulent Channel-Flow Dataset}
    \vspace{10pt} 
    \begin{tabular}{c  c  c  c  c  c  c   c  c   c  c  c  c   c   c    c} 
      $N_x$ && $N_y$ && $N_z$ && $Re_\tau$ & $dp/dx$ & $\nu$ & $u_*$ & $U_{bulk}$ && $\Delta x^+$ && $\Delta z^+$ & $\Delta t$\\
      \vspace{-8pt} \\
      2048 && 512 && 1536 && 1000 & $-2.5\times 10^{-3}$ & $5\times 10^{-5}$  & $5\times 10^{-2}$ & 1.00 && 12.3 && 6.1 & $1.3\times 10^{-3}$ \\
      \vspace{1pt} 
     \end{tabular}
  \end{center}
\end{table}

The JHTDB Web services supports subroutine-like calls for data that can be made from MATLAB, Fortran, and C/C++ programs, 
including {\it getVelocity} for stored velocity data and {\it getVelocityGradient} for finite-difference approximations to velocity-gradients; 
see \cite{li2008public,graham2016web}.  
To return data off the numerical space-time grid, the JHTDB Web services provide methods for performing spatial and temporal 
interpolation within the database. Spatial interpolation is supported using multivariate polynomial interpolation in the barycentric 
Lagrange form, with stencils containing q=4,6,8 points in each coordinate direction for an order q interpolant, so-called Lag4, Lag6, Lag8 
methods, respectively. Temporal interpolation is supported using cubic Hermite interpolating polynomials (PCHIPInt), with centered 
finite-difference evaluation of the end-point time derivatives, employing a total of four temporal points. Spatial differentiation at grid points is performed 
using differentiation matrices obtained from the barycentric method of Lagrange polynomial interpolation for $q=4,6,8$ points, 
so-called FD4NoInt, FD6NoInt, FD8NoInt options \citep{berrut2004barycentric,graham2016web}. The FD4Lag4 option provides 
off-grid space-interpolated gradients with derivatives approximated by the FD4NoInt method at the grid sites and these data then 
space-interpolated to queried points using the Lag4 method. In the current study, {\it getVelocity} is used with Lag6 and PCHIPInt options
to obtain off-grid velocity data and {\it getVelocityGradient} is used with FD4Lag4 and PCHIPInt options to obtain velocity-gradients. 

Because the JHTDB turbulent channel-flow dataset was obtained from a numerical simulation with relatively
low near-wall resolutions in the streamwise and spanwise directions (see Table 1), the vorticity 
$\bom(\bx,t)$ returned by the database in the vicinity of the wall may exhibit observable departures 
from an exact Navier-Stokes solution. The stochastic Cauchy invariant approximated by equations 
\eqref{eq3-7}, \eqref{eq3-9} thus cannot be expected to be exactly conserved in time using such numerical 
data, even if the stochastic Lagrangian integration scheme has $\Delta s$ sufficiently small and $N$ sufficiently 
large to be very well converged, since discretization errors depending upon $\Delta x,$ $\Delta y,$ $\Delta z,$ $\Delta t$ 
will remain. 
On the other hand,  a {\it coarse-grained vorticity} field 
\be \widehat{\bom}(\bx,t) = \int d^3x \ G_{\ell_x}(x-x')G_{\ell_y}(y-y')G_{\ell_z}(z-z')\ \bom(\bx',t) \lb{eq3-11} \ee 
when calculated from the database, with length-scales $\ell_x,$ $\ell_y,$ $\ell_z$ each comprising several grid-lengths
$\Delta x,$ $\Delta y,$ $\Delta z,$ will plausibly agree with a Navier-Stokes solution coarse-grained 
over the same length-scales, with better correspondence than for pointwise fields. Of course, 
the coarse-grained stochastic Cauchy invariants for a Navier-Stokes solution
\be \widehat{\tilde{\bom}}_s(\bx,t) = \int d^3x \ G_{\ell_x}(x-x')G_{\ell_y}(y-y')G_{\ell_z}(z-z')\ \tilde{\bom}_s(\bx',t) \lb{eq3-12} \ee 
will satisfy a corresponding statistical conservation law
\be \widehat{\bom}(\bx,t) = \bE\big[\widehat{\tilde{\bom}}_s(\bx,t) \big] \lb{eq3-13} \ee 
as a direct consequence of the conservation of the fine-grained invariants. If the conjecture is correct that 
the coarse-grained numerical solutions will agree well with a coarse-grained exact Navier-Stokes solution, then 
we may expect to observe better conservation numerically for the coarse-grained invariants. To test this 
hypothesis in our study below, we shall use box/tophat filters with $\ell_i=n_i (\Delta x_i)$ for integers $n_i,$ 
$i=x,y,z.$ In practice, this means that we locate the grid-point $\bx_k$ that lies closest to a given point $\bx$
and then average over the set of points $\bx'$ in the box centered at $\bx_k$ and extending $n_i/2$ 
grid spacings to the left and to the right in each coordinate direction $i=x,y,z.$ This space-average is easily 
included in our Monte Carlo scheme, by releasing stochastic Lagrangian particles at time $t$ 
with positions $\bx'$ chosen at random, uniformly distributed over such a box.

\subsection{Validation of the Numerical Method}\lb{sec:valid}  

We shall now validate the computational algorithms described above by verifying the 
mean conservation law of the stochastic Cauchy invariant, Eq.(\ref{eq2-50}). This is 
a stringent test of our numerical methods to evaluate the Cauchy invariant and its statistics 
and also of the fidelity of the JHTDB channel-flow dataset to a true Navier-Stokes solution.

\begin{table} 
    \begin{center}
    \vspace{-10pt} 
    \caption{Coordinates of Analyzed Vorticity Vectors}
    \vspace{10pt} 
    \begin{tabular}{c  c  c  c  c  c  c  c  c} 
                      &&  $x$  &  $y$  &  $z$  &  $t$ & $\omega_x$  & $\omega_y$  & $\omega_z$ \\
      ejection  &&  21.094707  &  0.994647 & 7.563944  & 25.9935 & -2.24948978 & -0.110395804 &  22.1811829\\
      \vspace{-8pt} \\
      sweep  &&  0.715000 & 0.995100  &  0.725900  & 25.9935 &  0.05745593 &  -0.1597188  &  47.2303467\\
      \vspace{1pt} 
     \end{tabular}
  \end{center}
   \label{tab:table2}
\end{table}

\subsubsection{Description of the Test Events and Variables}

Before presenting 
numerical results, we must recall that the stochastic Cauchy invariant is a {\it vector} quantity,
each of whose components $\tilde{\omega}_{s i}(\bx,t)$ for $i=x,$ $y,$ $z$ is conserved on average, 
with mean value equal to $\omega_i(\bx,t)$ independent of $s<t.$ A coordinate-free decomposition 
of the stochastic Cauchy invariant can be obtained using the unit vector $\bn_\omega(\bx,t):=\bom(\bx,t)/|\bom(\bx,t)|$ 
pointing in the direction of vorticity at final time $t,$ which permits definition of parallel and perpendicular components 
of the Cauchy invariant 
\be \widetilde{\omega}_{s\|}(\bx,t):=\widetilde{\bom}_{s}(\bx,t)\cdot \bn_\omega(\bx,t), \quad 
\widetilde{\bom}_{s\perp}(\bx,t):=\widetilde{\bom}_{s}(\bx,t)-\widetilde{\omega}_{s\|}(\bx,t) \bn_\omega(\bx,t),
\lb{eq5-1} \ee
which satisfy the mean conservation laws 
\be \bE\left[ \widetilde{\omega}_{s\|}(\bx,t)\right] = |\bom(\bx,t)|, \quad 
\bE\left[ \widetilde{\bom}_{s\perp}(\bx,t)\right] =\bzed, \qquad s<t.  \lb{eq5-2} \ee
In our numerical study below we shall present results both for the Cartesian components of 
the stochastic Cauchy invariant and also for the intrinsic $\|,$ $\perp$ components 

We have selected for investigation from the channel-flow database two test vorticity 
vectors $\bom(\bx,t)$, whose detailed space-time coordinates are given in Table 2. 
The choice of these two cases is discussed in detail in paper II. Here it suffices to note 
that the spatial coordinates both correspond to points at the bottom of the buffer-layer, with $y^+\simeq 5,$
one at the location of an ``ejection'' where fluid is erupting away from the wall and the 
other at a "sweep" where fluid is splatting against the wall. We shall refer to these two
events hereafter simply as the ``ejection'' and the ``sweep'', respectively.

\subsubsection{Statistics of the Cauchy Invariant for the Ejection Event}

\begin{figure}
\hspace{-50pt} 
\begin{subfigure}[b]{.6\textwidth}
  \centering
  \includegraphics[width=1.1\linewidth]{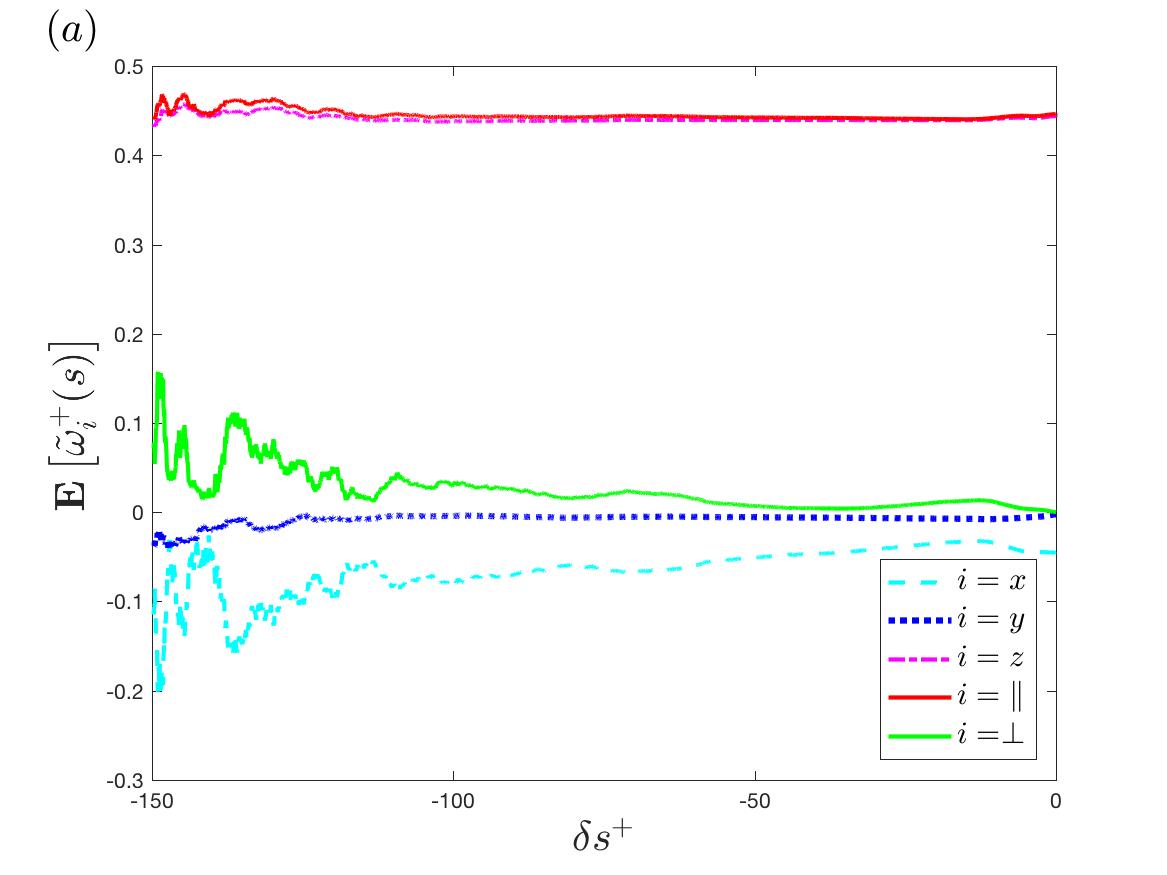}  
\end{subfigure}
\begin{subfigure}[b]{.6\textwidth}
  \centering
  \includegraphics[width=1.1\linewidth]{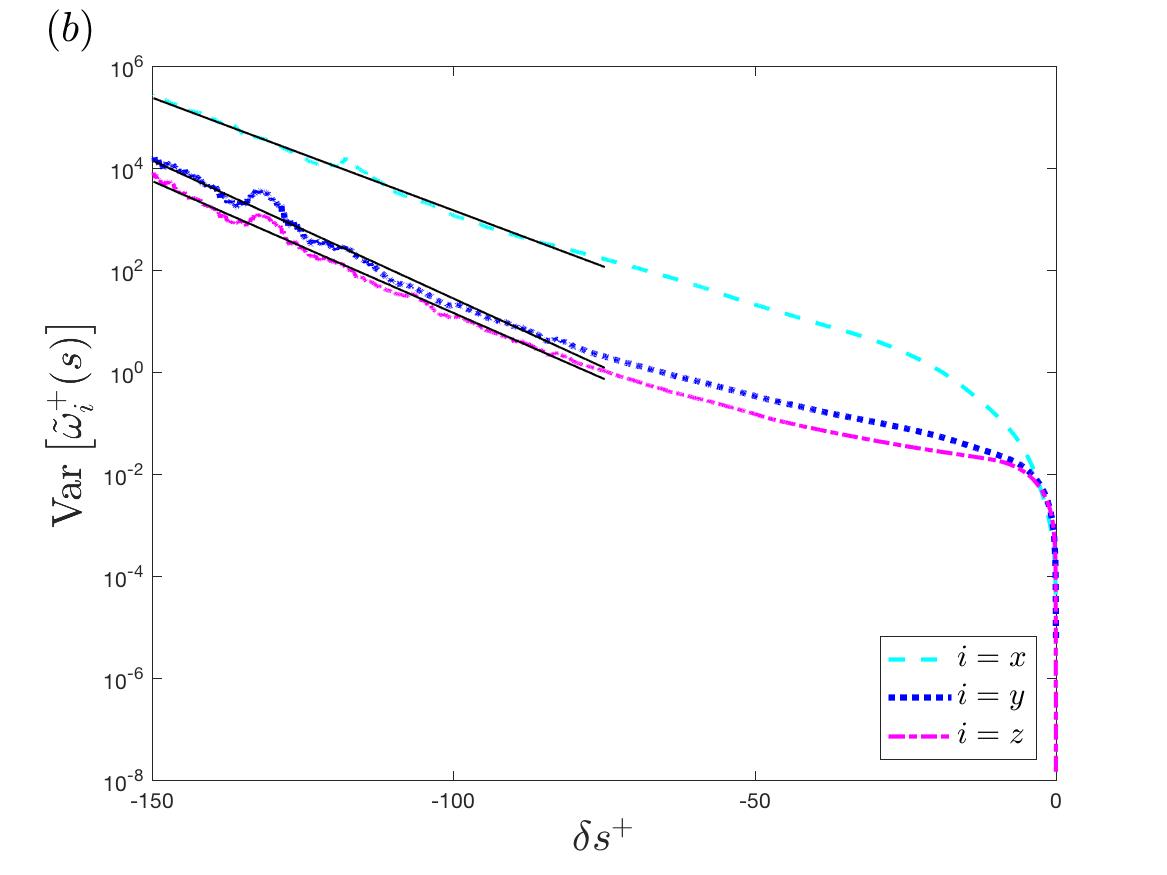}  
\end{subfigure}
\caption{Expectation value (a) and variance (b) of the components $\tilde{\omega}_{s i}^+(\bx,t)$ 
of the stochastic Cauchy invariant for the selected point $(\bx,t)$ in the ejection event, 
plotted as functions of $\delta s^+$,  calculated numerically with $\Delta  s=10^{-3}$ and $N=10^7.$ The thin
black lines are linear fits in semilogarithmic coordinates for $\delta s^+<-75.$} 
\lb{fig5-cauchyL}
\end{figure}

We first consider the ejection case. 
In Fig.~\ref{fig5-cauchyL}(a) we plot as functions of $\delta s=s-t$ the numerically calculated expectation values of the $i$th 
component of the stochastic Cauchy invariant for $i=x,$ $y,$ $z,$ $\|,$ and the vector norm $|\,\bE\left[ \widetilde{\bom}_{s\perp}(\bx,t)\right]|$
for $i=\perp.$ These were obtained with the numerical Monte Carlo scheme in Section \ref{sec:num}, 
using integration time-step $\Delta s=10^{-3}$ in Eqs.\eqref{eq3-1},\eqref{eq3-4} and number of particles 
$N=10^7$ to calculate sample averages as in Eq.\eqref{eq3-10}. 
According to the mathematical theory for exact Navier-Stokes solutions presented in Section \ref{sec:NS}, the 
graphs for all components in Fig.~\ref{fig5-cauchyL}(a) should be flat horizontal lines, with values equal 
to the corresponding components of the vorticity vector $\bom(\bx,t)$ at the final time. While this is reasonably 
verified at $\delta s^+>-100$, especially for components $i=y,z,\|,$ there are clear deviations at longer times and 
for other components $i=x,\perp.$ It should be emphasized that the ``naive Cauchy invariants'', which are exactly 
conserved for smooth Euler solutions, exhibit much larger deviations from conservation for the channel-flow 
Navier-Stokes solution than do the numerical results in Fig.~\ref{fig5-cauchyL}(a) for the stochastic Cauchy 
invariants.  As discussed in the SM, the ``naive invariants'' can be accurately evaluated by the same numerical 
methods as discussed in Section \ref{sec:num}, by setting the random noise to zero and calculating 
a standard (deterministic) Lagrangian trajectory and the deformation matrix along it by a 4th-order 
Runge-Kutta scheme. The results in the SM show that conservation of the ``naive invariants'' is 
violated by several orders of magnitude for the same case as in Fig.~\ref{fig5-cauchyL}(a). 
There is, of course, no reason that conservation of the standard Cauchy invariants should 
hold for any Navier-Stokes solution and the failure of conservation here confirms previous results 
that vorticity  is not even approximately ``frozen in'' for a high Reynolds-number turbulent flow 
\citep{luthi2005lagrangian,guala2005evolution,guala2006stretching,johnson2016large,johnson2017analysis}.
Nevertheless, despite the much better conservation of the stochastic Cauchy invariants that we observe, 
the deviations from exact conservation that arise in our various numerical approximations must be clearly 
understood. 

A first possible source of error is the size of the time step. The value of $\Delta s=10^{-3}$ employed 
for stochastic integration was slightly smaller than the time-step $\Delta t$ employed in the original channel-flow 
simulation archived in the database (see Table 1).  Furthermore, as discussed in the SM, we checked that the 
numerical results with $\Delta s=10^{-3}$ are essentially unchanged when using a twice smaller value 
$\Delta s=5\times 10^{-4}$ over the reduced time range $-100<\delta s^+<0.$ The average difference over 
that interval between the values plotted in Fig.~\ref{fig5-cauchyL}(a) and those generated with $\Delta s=5\times 10^{-4}$   
is less than 0.004 for all components. In particular, the average differences for the parallel and perpendicular
components over that interval are, respectively, only 0.21\% and 0.79\% of the vorticity magnitude $|\bom|.$
Thus, our results appear to be well-converged in the time-step. 

A much larger source of numerical error in the results presented in Fig.~\ref{fig5-cauchyL}(a)
is the number of samples $N$ employed to calculate averages over realizations of the Brownian motion. 
The error in approximating expectation values from $N$-sample averages as in Eq.\eqref{eq3-10} is 
estimated from the Central Limit Theorem on order of magnitude as 
\be         \delta  \bE\left[ \widetilde{\omega}_{s\,i}(\bx,t)\right] \sim \sqrt{\frac{{\rm Var}[\widetilde{\omega}_{s\,i}(\bx,t)]}{N}}.  
\lb{eq5-3} \ee
We can numerically calculate the variances within the same Monte Carlo scheme by using the standard 
unbiased estimator 
\be {\rm Var}[\tilde{F}] \doteq \frac{1}{N-1} \sum_{n=1}^N (\tilde{F}^{(n)}-{\mathbb E}[\tilde{F}])^2, \lb{eq5-4} \ee
and variances of the Cartesian coordinates of the stochastic Cauchy invariant estimated in this manner 
with $\Delta s=10^{-3},$ $N=10^{7}$ are plotted in Fig.~\ref{fig5-cauchyL}(b) versus $\delta s^+.$ After 
an initial transient period, these variances appear to grow exponentially rapidly backward in time, 
with ${\rm Var}[\widetilde{\omega}_{s\,i}(\bx,t)]\propto \exp(2\lambda_i(t-s))$ for $s\ll t.$ The growth exponents
obtained by linear fits of the numerical results over the range $\delta s^+<-75 $ in semilogarithmic coordinates are 
\be  \lambda_x^+=0.0509,  \quad  \lambda_y^+=0.0624,  \quad  \lambda_z^+=0.0596.  \lb{eq5-5} \ee
in wall units. Recall that, according to Eq.~\eqref{eq2-46}, the stochastic Cauchy invariant evolves in the same 
manner as does a material line element forward in time, except that the motion is along stochastic 
Lagrangian trajectories rather than deterministic ones. The exponential growth of variances that we 
find is thus consistent with the previous study of \cite{johnson2017analysis}, who observed Lagrangian
chaos in the same channel-flow database that we employ. The instantaneous Lyapunov exponents 
obtained by \cite{johnson2017analysis} are plotted in their Figure 4(b), with values $2\times 10^{-3}\sim 2\times 
10^{-2}$ in the range $0<y^+<10$ where their non-dimensionalization by a local Kolmogorov time (increasing with $y$)
corresponds essentially to our wall units. Our growth exponents for the variance are order of magnitude consistent
with those Lyapunov exponents, but somewhat larger. The somewhat greater values could be due to the 
fact that  we do not average over long times, as \cite{johnson2017analysis} did, and the local stretching 
rates in our extreme-stress event could be larger than average. Also, we do not conditionally sample on a 
given $y^+$-value as \cite{johnson2017analysis} did, and the stochastic Lagrangian trajectories released at 
$y^+=5.35$ can disperse into the buffer-layer where local stretching rates are highest \citep{johnson2017analysis}.
 
The errors in the mean values estimated from the Central Limit Theorem, using Eq.~\eqref{eq5-3} 
and the variance calculated by Eq.~\eqref{eq5-4}, explain the largest deviations from 
exact conservation observed in Fig.~\ref{fig5-cauchyL}(a). For $\delta s^+<-100,$ 
all of the observed deviations are consistent on order of magnitude with fluctuations due 
to finite $N.$ Because of the slow Monte Carlo rate of convergence $\propto 1/\sqrt{N}$
in Eq.~\eqref{eq5-3}, it would require $N\simeq 10^{14}$ samples to achieve accuracy of even
a few percent for $\delta s^+\simeq -150,$ which is beyond our computational means. On the other hand, 
we estimate from Eq.~\eqref{eq5-3} that Monte Carlo errors are much less than 1\% for  $\delta s^+>-100.$
This is verified by the observed convergence of the mean values with increasing $N$ (not shown) 
and the smoothness of the plotted curves for $\delta s^+>-100.$ In that range, we see that the 
components $i=y,z,\|$ of the stochastic Cauchy invariant are well conserved in the mean, 
but there are observable deviations from conservation for $i=x,\perp.$ In fact, the mean
deviation of the parallel and perpendicular components from conservation over the range 
$-100<\delta s^+<0$ are 0.57\% and 2.97\% of the magnitude $|\bom|,$ respectively. 
The final-time vorticity direction vector is nearly $\bn_\omega\simeq \hat{{\bf z}}$ for
the case being examined, so that the error in conservation of the perpendicular component
arises mainly from non-conservation of the $x$-component. These violations of exact 
conservation are not explained by any numerical artifacts of finite $\Delta s$ 
and $N$ in our Monte Carlo integration scheme. 

\begin{figure}
\hspace{-50pt} 
\begin{subfigure}[b]{.6\textwidth}
  \centering
  \includegraphics[width=1.1\linewidth]{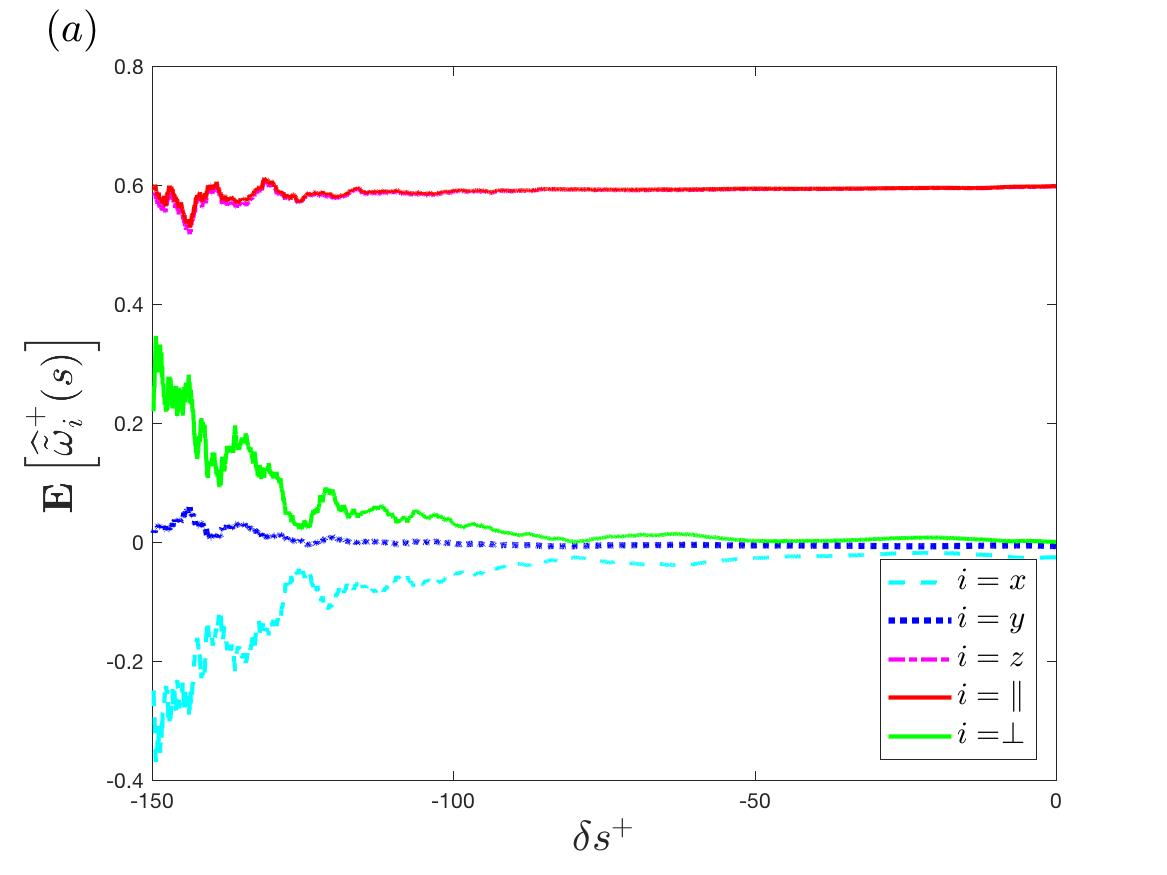}  
\end{subfigure}
\begin{subfigure}[b]{.6\textwidth}
  \centering
  \includegraphics[width=1.1\linewidth]{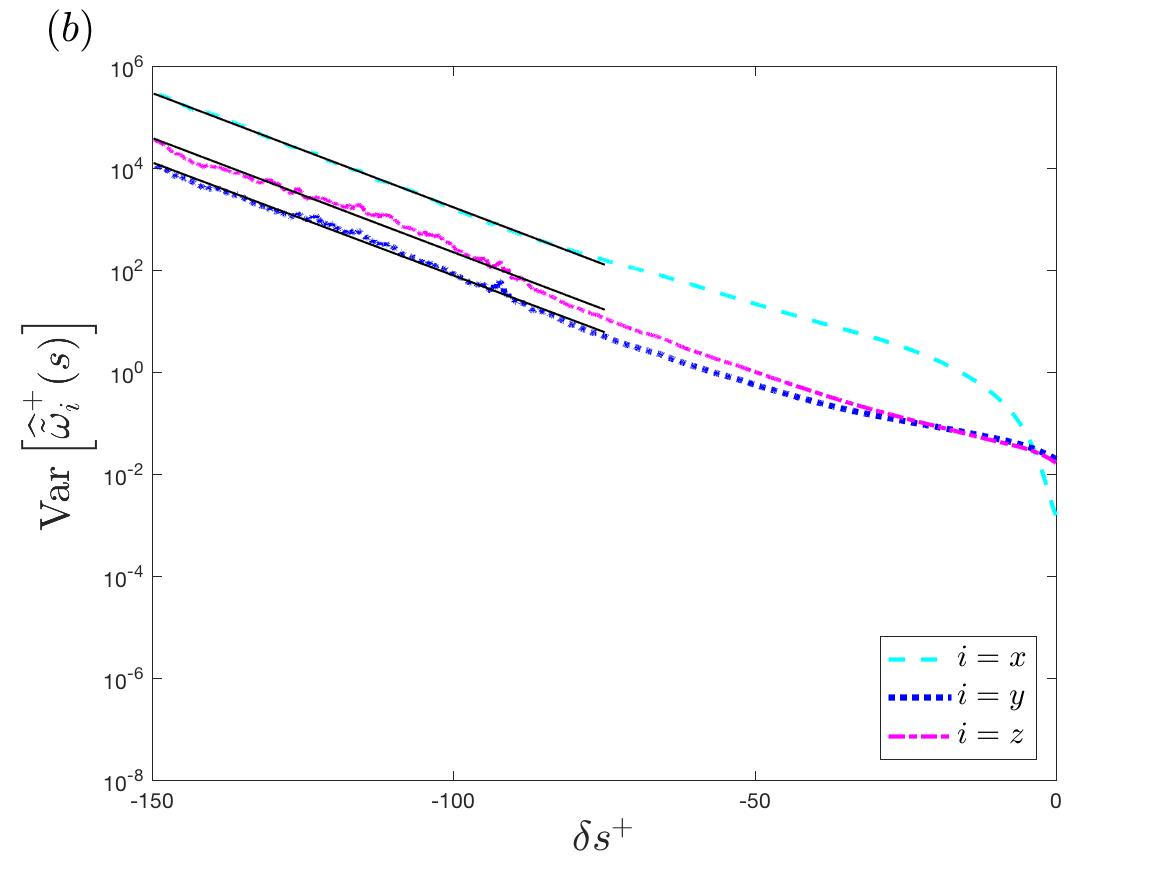}  
\end{subfigure}
\caption{Expectation value (a) and variance (b) of the components $\widehat{\tilde{\omega}}_{s i}^+(\bx,t)$ 
of the coarse-grained Cauchy invariant for the selected point $(\bx,t)$ in the ejection event, 
plotted as functions of $\delta s^+$,  calculated numerically with $\Delta  s=10^{-3}$ and $N=10^7.$ The thin
black lines are linear fits in semilogarithmic coordinates for $\delta s^+<-75.$} 
\lb{fig5-fcauchyL}
\end{figure} 

The only remaining source of error in our computation is the deviation of the fields stored 
in the JHTDB channel-flow database from an exact Navier-Stokes solution.
Although the numerical simulation stored in the database had excellent spatial resolution 
in the $y$-direction near the wall, the wall-parallel resolution was much poorer. As reviewed in 
Table 1, $\Delta z^+=6.1$ and $\Delta x^+=12.3.$ Furthermore, the velocity-gradients returned 
by the database function {\it getVelocityGradient} are finite-difference approximations and 
both velocities and velocity-gradients at points off the computational grid are supplied 
by Lagrangian interpolation. A reasonable estimate of the size of the errors involved in 
these approximations can be obtained from the deviation of the database velocity-gradient
matrices from being traceless, as required by flow incompressibility. As we discuss in 
the SM, we have found that the divergence-free condition is satisfied to a few
percent, with violation less than 0.5\% for $y^+<10$ but rising up to 3-4\% for $y^+>30.$
This inaccuracy in the returned velocity-gradients accords well with the magnitude of the 
deviations that we observe in mean conservation of the stochastic Cauchy invariant. 
An ensemble of stochastic Lagrangian particles  
released at $y^+=5.35$ will sample not a resolved Navier-Stokes solution but instead a Lagrange 
interpolating polynomial, until the ensemble has dispersed over at least several grid-lengths in the wall-parallel 
directions. This is the presumed source of the slight non-conservation observed for $i=x,\perp$ components 
at times $\delta s^+>-100$ in in Fig.~\ref{fig5-cauchyL}(a). To test this explanation, we have numerically 
generated the coarse-grained stochastic Cauchy invariant defined in Eq.~\eqref{eq3-12} for filter lengths
$\ell_i= n_i (\Delta x_i)$ in the coordinate directions, $i=x$, $y$, $z,$ using the Monte Carlo
scheme discussed in section \ref{sec:time-int}. As discussed in paper II, $(n_x,n_y,n_z)=(4,0,4)$ 
is a reasonable choice of filtering lengths which does not obscure the essential physics of the 
ejection event under consideration. 

Plotted in Fig.~\ref{fig5-fcauchyL} are the mean and variance of the coarse-grained stochastic
Cauchy invariant for the selected point, filtered over $(n_x,n_y,n_z)=(4,0,4)$ grid points. 
The fluctuations of the mean values for $\delta s^+<-100$ are increased compared with the 
unfiltered case shown in Fig.~\ref{fig5-cauchyL}(a), because the empirical average over  $N$ samples
is now doing double duty to represent  both the expectation over the Brownian motion and the 
space-average over the filtering box. As can be seen from Fig.~\ref{fig5-fcauchyL}(b), the variances 
of the coarse-grained Cauchy invariant are somewhat increased relative to those plotted in Fig.~\ref{fig5-cauchyL}(b),
because the former include both stochastic and spatial fluctuations. On the other hand, as expected, 
the mean-conservation of the coarse-grained Cauchy invariant is observed in Fig.~\ref{fig5-fcauchyL}(a)  
to be improved for $\delta s^+>-100$ when compared with the fine-grained means plotted in  
Fig.~\ref{fig5-cauchyL}(a), especially for the $i=x,$ $\perp$ components. The average deviation from 
conservation of the parallel and perpendicular components is now found to be 0.79\% and 1.22\%
of the magnitude $\left|\widehat{\bom}\right|,$ respectively. We have found that 
increasing $n_x,$ $n_z$ further (not shown) improves the mean conservation, while the spatial 
structures become more diffuse. Since mean conservation of the stochastic Cauchy invariant is 
a quite stringent test of validity of a Navier-Stokes solution, these results validate both our Monte Carlo 
numerical method to calculate the stochastic Cauchy invariant and also the adequacy of the 
JHTDB channel-flow database to investigate the turbulent buffer layer. 


\subsubsection{Statistics of the Cauchy Invariant for the Sweep Event}

\begin{figure}
\hspace{-50pt} 
\begin{subfigure}[b]{.6\textwidth}
  \centering
  \includegraphics[width=1.1\linewidth]{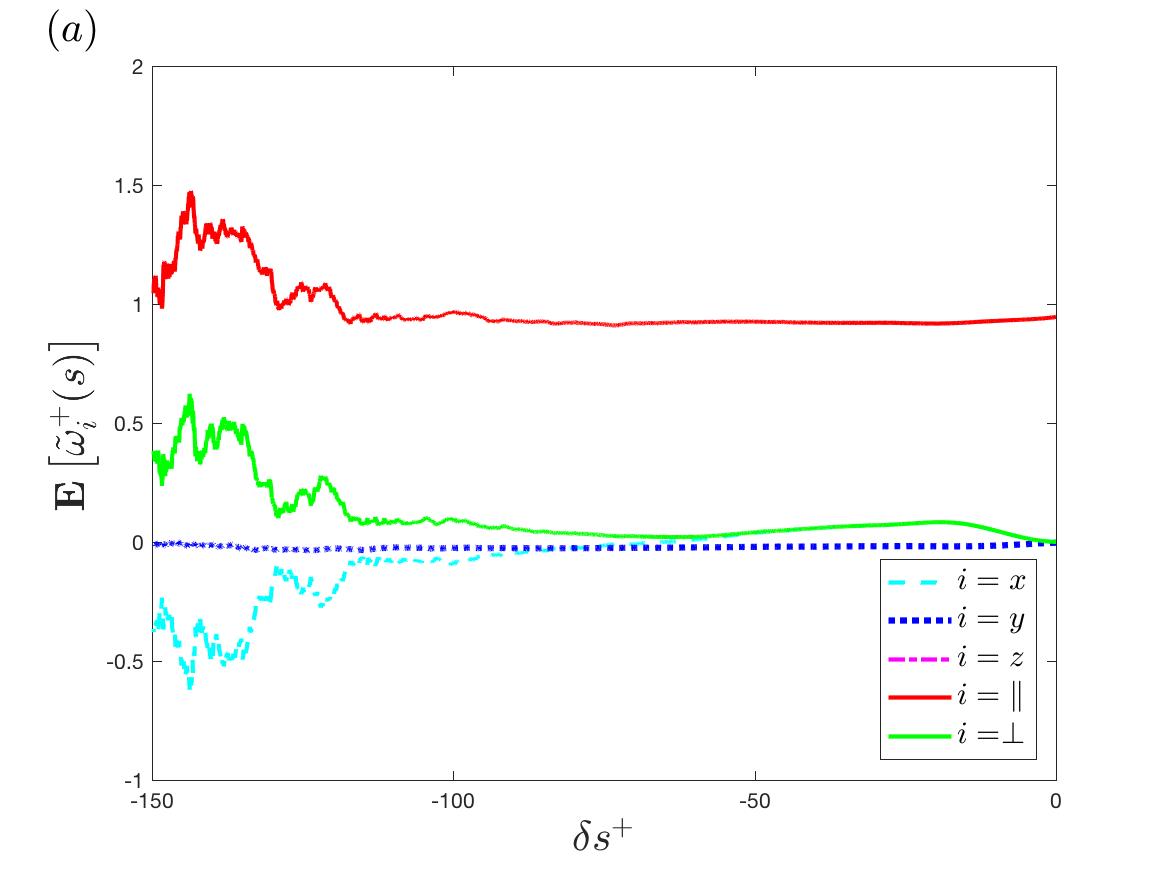}  
\end{subfigure}
\begin{subfigure}[b]{.6\textwidth}
  \centering
  \includegraphics[width=1.1\linewidth]{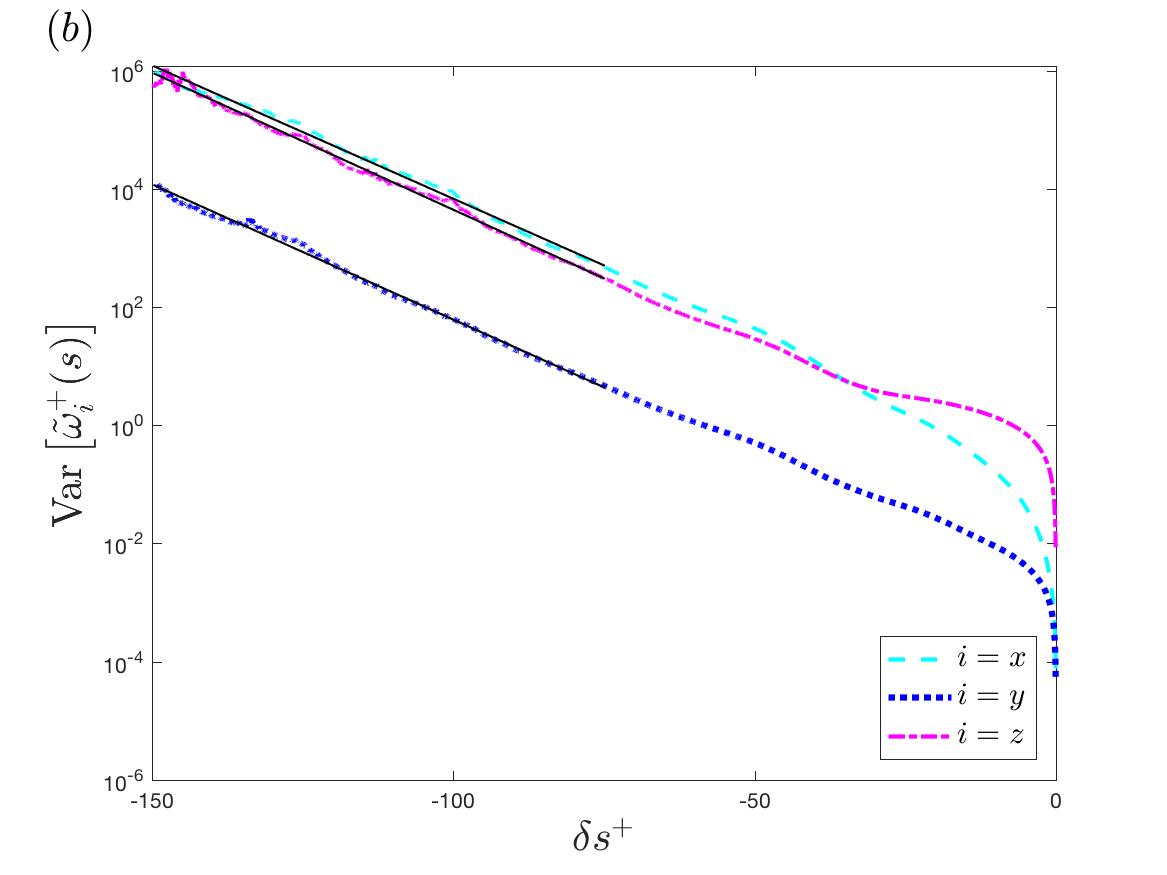}  
\end{subfigure}
\caption{Expectation value (a) and variance (b) of the components $\tilde{\omega}_{s i}^+(\bx,t)$ 
of the stochastic Cauchy invariant for the selected point $(\bx,t)$ in the sweep event, 
plotted as functions of $\delta s^+$,  calculated numerically with $\Delta  s=10^{-3}$ and $N=10^7.$ The thin
black lines are linear fits in semilogarithmic coordinates for $\delta s^+<-75.$} 
\lb{fig5-cauchyH}
\end{figure}

We now consider, more briefly, the sweep event.   We again investigate the mean conservation of the stochastic 
Cauchy invariant  as a check on our numerical approximations. Fig.~\ref{fig5-cauchyH} plots the results 
for the mean and the variance of the components of the Cauchy invariant, calculated by our Monte Carlo 
integration scheme with time-step $\Delta  s=10^{-3}$ and with number of samples $N=10^7.$ As for the 
ejection event, we have performed a convergence analysis in $\Delta s$ (see SM) and found 
that the results do not change significantly with a smaller time-step $\Delta  s=5\times 10^{-4}.$ 
Over the reduced interval $-100<\delta s^+<0$ the mean differences in results for the two time steps
differ by less than 0.005 for all components and the mean differences for the parallel and perpendicular 
components are only 0.38\% and 0.45\% of the magnitude $|\bom|,$ respectively. Due to the finite values of $N,$ 
however, the results for $\delta s^+<-100$ exhibit very large errors, consistent with the variances 
plotted in Fig.~\ref{fig5-cauchyH}(b). The results for  $-100<\delta s^+<0$ on the other hand are 
well-converged in $N,$ so that any residual errors there are due to lack of resolution in the archived channel-flow 
simulation. Mean conservation holds quite well for the $y$, $z$ components of the stochastic Cauchy invariant
in the time interval $-100<\delta s^+<0$ for this sweep event, but the $x$-component is less well conserved. 
Quantitatively, the mean deviations from conservation of the parallel and perpendicular components are
2.29\% and 5.01\% of the vorticity magnitude $|\bom|,$ respectively. 
We have again found that spatial coarse-graining noticeably improves mean conservation, 
although perhaps less well than it did in the ejection case. The mean deviations from conservation 
of the parallel and perpendicular components are now 
1.43\% and 3.32\% of the vorticity magnitude $\left|\widehat{\bom}\right|,$ respectively. 
We present plots of the coarse-grained stochastic Cauchy invariant for the sweep case in 
Fig.~\ref{fig5-fcauchyH}, where improved conservation can be observed especially for $-50<\delta s^+<0.$

\begin{figure}
\hspace{-50pt} 
\begin{subfigure}[b]{.6\textwidth}
  \centering
  \includegraphics[width=1.1\linewidth]{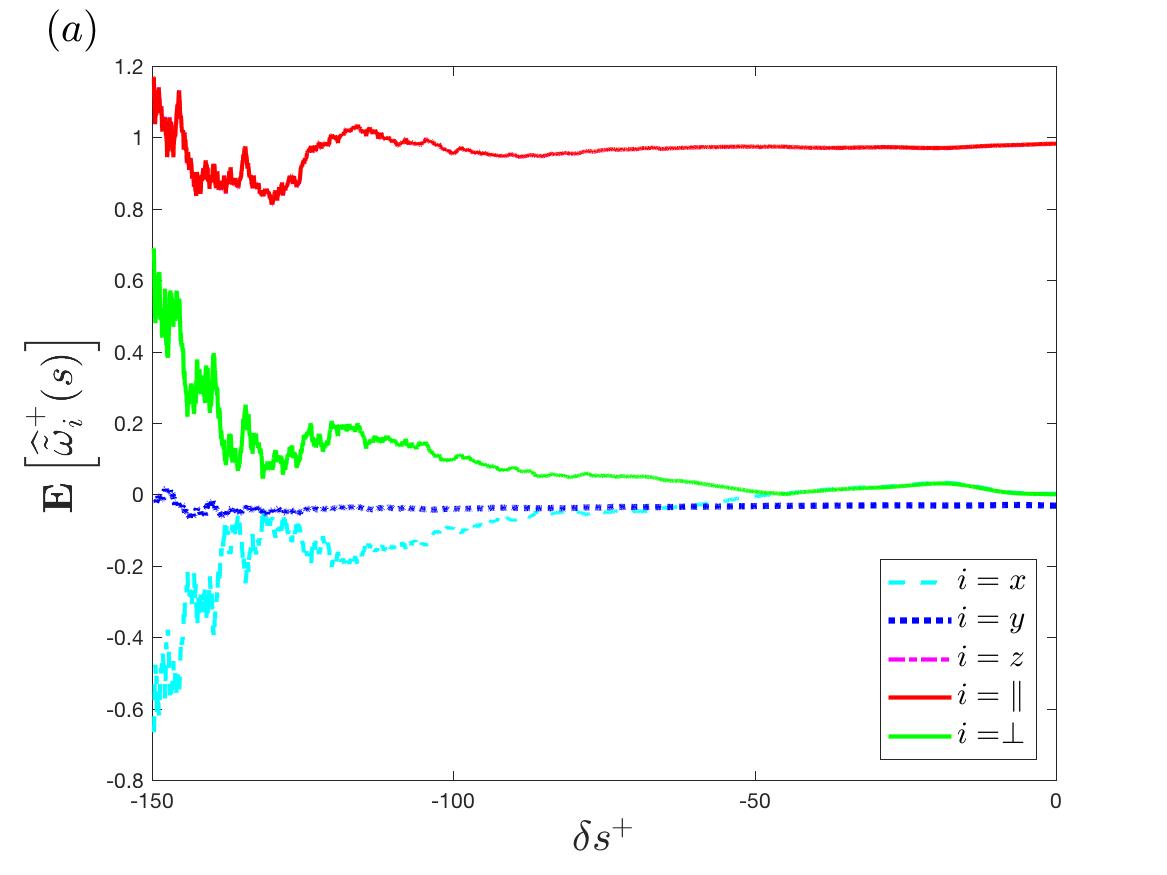}  
\end{subfigure}
\begin{subfigure}[b]{.6\textwidth}
  \centering
  \includegraphics[width=1.1\linewidth]{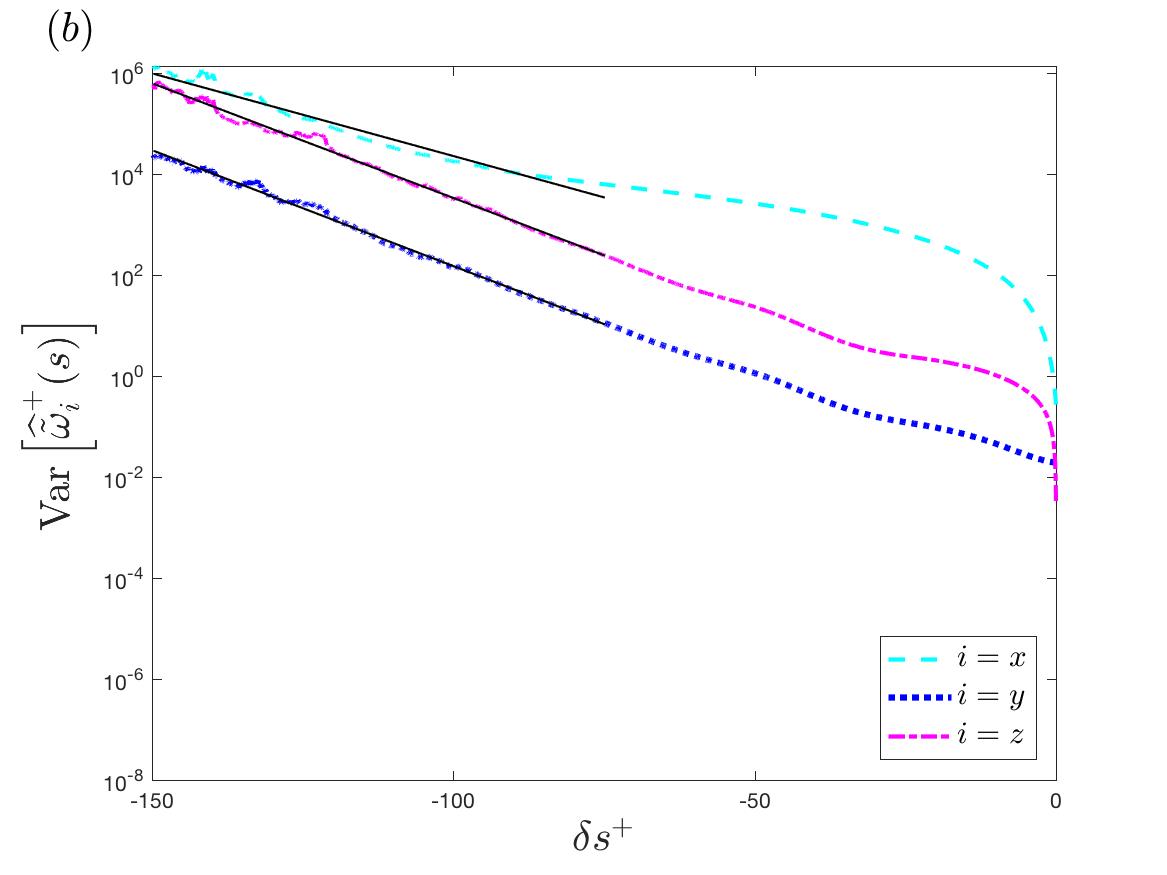}  
\end{subfigure}
\caption{Expectation value (a) and variance (b) of the components $\widehat{\tilde{\omega}}_{s i}^+(\bx,t)$ 
of the coarse-grained Cauchy invariant for the selected point $(\bx,t)$ in the sweep event, 
plotted as functions of $\delta s^+$,  calculated numerically with $\Delta  s=10^{-3}$ and $N=10^7.$ The thin
black lines are linear fits in semilogarithmic coordinates for $\delta s^+<-75.$} 
\lb{fig5-fcauchyH}
\end{figure}

The growth exponents for the variances in the sweep case obtained by linear fits over the range 
$\delta s^+<-75 $ in semilogarithmic coordinates are 
\be  \lambda_x^+=0.1039,  \quad  \lambda_y^+=0.1055,  \quad  \lambda_z^+=0.1069.  \lb{eq5-6} \ee
These are about double the exponents reported in \eqref{eq5-5} for the ejection event. The greater
size is reasonable, since in the sweep event the stochastic particles moving backward in time are 
lifted up by the flow higher into the buffer layer where local stretching rates are largest \citep{johnson2017analysis}.  
Just as for the ejection considered previously, we conclude that the stochastic Cauchy vector at time $s$ 
is exponentially stretched and rotated by transport from $s$ to $t,$ as pictured in the middle panel of 
Fig.~\ref{fig:CauchyFigure}. An important implication is that the constant mean of the stochastic Cauchy 
invariant demonstrated in Figs.\ref{fig5-cauchyL}(a),\ref{fig5-cauchyH}(a) is due to extensive cancellations 
of much larger magnitude vorticity vectors for individual realizations. We shall explore this finding 
further in paper II.  

\section{Conclusions and Prospects}\lb{sec:concl}  

We have explained the \cite{ConstantinIyer08,ConstantinIyer11} theory for Navier-Stokes solutions 
as a stochastic Lagrangian representation in the \cite{kuzmin1983ideal}-\cite{oseledets1989new} 
formulation of the equation. The theory yields exact mean conservation laws for 
stochastic Cauchy invariants (in the generalized sense of \cite{besse2017geometric}) 
corresponding both to vortex-momentum and to vorticity. For wall-bounded flows, these conservation 
laws yield an exact representation of vorticity at any interior point as an average over 
stochastic vorticity contributions transported from the wall. We discussed some relations 
of the \cite{ConstantinIyer08,ConstantinIyer11} results with the Eulerian theory of 
\cite{lighthill1963boundary}-\cite{morton1984generation} for generation of vorticity 
at solid walls. We have also elaborated a numerical Monte Carlo method to calculate the 
stochastic Cauchy invariants and their statistics, given the Eulerian flow fields in space-time. 
These results hold for every Navier-Stokes solution, laminar as well as turbulent. We believe 
that the theory justifies stochastic Lagrangian flows as natural for viscous hydrodynamics, 
providing deeper analysis and insights into nonlinear vortex dynamics than the deterministic Lagrangian 
trajectories more suited to ideal fluids. In particular, the stochastic formulation fully represents
the creation of vorticity at solid walls and its diffusion into the interior.  

In the following paper II, we shall apply these exact mathematical and computational tools 
to investigate a concrete problem, the process of ``lifting'' of vortex lines from the wall into the 
buffer layer of turbulent channel flow. The present paper has validated the numerical Lagrangian 
scheme proposed in this work when used in conjunction with the simulation database for channel-flow 
turbulence documented by \cite{graham2016web}. In particular, we have shown that the 
stochastic Cauchy invariant calculated by our methods has a well-conserved mean value,
providing a stringent test on the fidelity of the numerical results to an exact Navier-Stokes 
solution. In the following paper, we shall explore the implications of the large cancellations 
in the conserved means of the stochastic Cauchy invariants which is implied by the exponential 
growth in their variances. We shall compare there our results and conclusions
with those drawn from earlier experimental and computational studies. The long term 
goal of this line of research is to understand in detail the Lagrangian evolution of vorticity 
generated at the wall and to explicate the mechanisms by which a constant mean flux 
of spanwise vorticity is maintained.  

An important related direction for future work is more efficient numerical methods.  
In this paper we have developed and applied the simplest Monte Carlo scheme with an 
Euler-Maruyama discretization of the equations of motion for stochastic Lagrangian particles. Unfortunately, 
the convergence rate in number of samples is quite slow, which significantly limits the range of applications. 
Within standard Monte Carlo, importance sampling techniques, e.g. 
by change of drift \citep{arouna2004adaptative}, can provide variance reduction and more efficient sampling 
of rare events. To get fundamental improvement of Monte Carlo rates of convergence, however, alternatives  
such as quasi-Monte Carlo \citep{hofmann1997quasi} and multilevel Monte Carlo \citep{higham2013mean} 
should be explored. We believe that development of such improved numerical schemes is well 
worth the effort, since stochastic Lagrangian analysis provides unique information on the flow 
physics not accessible by other means. 

There are also clear directions for further work on mathematical and physical theory. As we shall discuss in 
\cite{eyink2020stochastic}, the stochastic Lagrangian approach of \cite{ConstantinIyer08,ConstantinIyer11} 
may be directly related to the Eulerian theory of \cite{lighthill1963boundary}-\cite{morton1984generation} 
for vorticity generation at the wall. Closer ties should also exist with the \cite{taylor1932transport}-\cite{huggins1994vortex}
relation, which continues the mean vorticity flux into the interior of the flow. Our discussion in section 
\ref{sec:LM} relates the stochastic Lagrangian trajectories to the Eulerian vorticity flux $\bSigma$ 
integrated over any closed, bounded surface, but this result holds with any possible choice of flux. 
For example, $\Sigma_{ij}^*:=\epsilon_{ijk} \dot{u}_k$ is a perfectly acceptable definition of a ``vorticity flux''
which satisfies the balance equation \eqref{eq3}, but its long-time average in the steady-state is zero!
It should be possible to measure the distinguished flux \eqref{eq2} directly with stochastic trajectories,
either pointwise or integrated over wall-parallel planes. If so, interesting physics questions could be addressed,
such as how fresh vorticity injected at the wall is transported subsequently throughout the flow. Without 
a detailed proof,  we note that an apparently natural measure of vorticity flux in terms of the stochastic 
Cauchy invariant can be calculated as
\be  \lim_{s\to t-} \frac{1}{t-s}\bE\left[(\bx-\tbA^s_t(\bx))\btimes \tilde{\bom}_s(\bx,t)\right]= 
\bu(\bx,t)\btimes \bom(\bx,t)-2\nu\grad\btimes\bom(\bx,t). \lb{eq6-1} \ee
The viscous contribution in \eqref{eq6-1} differs from that in the Eulerian flux \eqref{eq2} by a factor of 2, 
unfortunately, 
although this calculation does suffice to show that there must be organized vorticity transport by stochastic 
trajectories in the steady-state. Another theoretical topic deserving further work is the discrete approximation
of continuous flows suggested by both \cite{kuzmin1983ideal} and \cite{oseledets1989new} in terms of $N$ 
point-dipoles satisfying the \cite{roberts1972hamiltonian} equations. Fluid turbulence makes it natural to
consider the statistical mechanics of such dipole systems. Just as with point-vortices, the regime of interest
to statistical hydrodynamics is a non-extensive, high-energy limit with the flow volume fixed and the energy 
growing $\propto N^2$ \citep{eyink1993negative}.  Finally, stochastic action principles analogous to 
that of \cite{eyink2010stochastic} can be investigated within the \cite{kuzmin1983ideal}-\cite{oseledets1989new} 
framework, especially for wall-bounded flows. It would be interesting to know in that case whether the 
conservation laws \eqref{eq2-48}, \eqref{eq2-50} for stochastic Cauchy invariants can be derived as 
consequences of particle relabelling symmetry. 

\section*{Acknowledgements}

We are grateful to B\'ereng\`ere Dubrulle and Charles Meneveau for useful discussions and suggestions 
on this paper. We acknowledge the NSF grant BigData:OCE-1633124 for support and G.E. also
acknowledges the Simons Foundation through Targeted Grant in MPS-663054 for partial support. 
This research project was conducted using simulation data from the Johns Hopkins Turbulence 
Database (JHTDB) and scientific computing services at the Maryland Advanced Research Computing 
Center (MARCC).


 \appendix 
 
\section{Stochastic Interpolation for Hitting Times \& Locations}\label{sec:interp}

We consider first the stochastic interpolation for the hitting time $\tilde{s}_*$ based upon the wall-normal 
coordinate $\tilde{b}(s)$ of the stochastic particle. We suppose that the numerical integration has given 
$\tilde{b}(s_i)=b_i$, $i=0,1$ for times $s_1<s_0$ with $|b_0|<h$ and $|b_1|>h.$ For a particle hitting 
the wall at $b=\pm h,$ the change of coordinates $b\to b'=h \mp b$ transforms the crossing into 
the event $\tilde{b}'(s_0)=b_0'>0$ and $\tilde{b}'(s_1)=b_1'<0.$ Hereafter we assume this transformation 
has been made and drop the primes. For small $\Delta s=s_0-s_1,$ the particle process can be written 
at times $s\in (s_1,s_0)$ as  
\be \tilde{b}(s) = b_0 + v(s-s_0) -\sqrt{2\nu}\, \tilde{W}_2(s), \lb{eqA-1} \ee
where $\tilde{W}_2(s)$ for $s<s_0$ is a time-reversed Brownian motion with $\tilde{W}_2(s_0)=0.$
Imposing the conditions $\tilde{b}(s_0)=b_0>0$ and $\tilde{b}(s_1)=b_1<0$ then 
implies that $\tilde{W}_2(s)$ is a generalized {\it Brownian bridge} satisfying for $\Delta b:=b_0-b_1>b_0>0$
\be  \tilde{W}_2(s_0)=0<\frac{b_0}{\sqrt{2\nu}} , \quad \tilde{W}_2(s_1)=\frac{\Delta b - v\Delta s}{\sqrt{2\nu}} 
> \frac{b_0 - v\Delta s}{\sqrt{2\nu}}. \lb{eqA-2} \ee
The first hitting time $\tilde{s}_*\in (s_1,s_0)$ when $\tilde{b}(\tilde{s}_*)=0$ then corresponds to the largest 
crossing time of the straight line 
$b=(b_0+v(s-s_0))/\sqrt{2\nu}$ and the Brownian bridge $\tilde{W}_2(s)$, or  
\be \tilde{W}_2(s) = \frac{b_0+v(s-s_0)}{\sqrt{2\nu}} \lb{eqA-3} \ee

Determining this hitting time $\tilde{s}_*$ can be mapped, by a sequence of transformations, onto a solved problem 
of determining the first hitting time of a standard Brownian motion with a straight line.  First, $\tilde{W}_2(s)$ can be 
related to a standard Brownian bridge $\tilde{B}^{(0)}_2(s),$ satisfying
\be \tilde{B}^{(0)}_2(s_0) =0, \quad \tilde{B}^{(0)}_2(s_1)=0, \lb{eqA-4} \ee
by the transformation 
\be \tilde{W}_2(s) = \left[\frac{\Delta y - v\Delta s}{\sqrt{2\nu}}\right] \frac{s-s_0}{s_1-s_0} + \tilde{B}^{(0)}_2(s). 
\lb{eqA-5} \ee
See \cite{borodin2015handbook}, Part I, Chapter IV.4.21 (p.64, final statement). 
The first hitting time $\tilde{s}_*$ is then determined as the largest time $s$ such that 
\be \tilde{B}^{(0)}_2(s) = \frac{1}{\sqrt{2\nu}} \frac{b_0(s_1-s)+b_1(s-s_0)}{s_1-s_0}. \lb{eqA-6} \ee
Next, the standard Brownian bridge $\tilde{B}^{(0)}_2(s)$ can be mapped onto a standard Brownian motion
$\tilde{B}_2(\sigma)$ via the transformation 
\be  \tilde{B}^{(0)}_2(s) = \frac{s_1-s}{\sqrt{\Delta s}} \tilde{B}_2\left(\frac{s-s_0}{s_1-s}\right) \lb{eqA-7} \ee
with the non-negative, dimensionless time variable $\sigma:=\frac{s-s_0}{s_1-s}\in [0,\infty)$. 
See \cite{borodin2015handbook}, Part I, Chapter IV.4.21(b) (p.64). The first-hitting time is then given 
by the inverse formula $\tilde{s}_*=\frac{s_0+\tilde{\sigma}_*s_1}{1+\tilde{\sigma}_*},$ 
where $\tilde{\sigma}_*$ is the smallest time $\sigma$ such that  
\be \tilde{B}_2(\sigma) = \frac{b_0+b_1\sigma}{\sqrt{2\nu\Delta s}} =\alpha-\beta\sigma, \qquad
\alpha = \frac{b_0}{\sqrt{2\nu\Delta s}}, \quad \beta=\frac{|b_1|}{\sqrt{2\nu\Delta s}} \lb{eqA-8} \ee
Equivalently, $\tilde{s}_*=s_0-\Delta \tilde{s}_*$ with $\Delta \tilde{s}_*=\Delta s \frac{\tilde{\sigma}_*}{1+\tilde{\sigma}_*}.$

The statistical distribution of the first-hitting time $\tilde{\sigma}_*$ of a standard Brownian motion $\tilde{B}_2(\sigma)$
and the straight line $\alpha-\beta\sigma$ is well-known to have the form 
\be p(\sigma) = \frac{|\alpha|}{\sqrt{2\pi \sigma^3}}\exp\left( - \frac{(\alpha-\beta\sigma)^2}{2\sigma}\right) 
= \left(\frac{\lambda}{2\pi \sigma^3}\right)^{1/2} \exp\left( - \frac{\lambda(\sigma-\mu)^2}{2\mu^2\sigma}\right) \lb{eqA-9} \ee 
with parameters
\be \lambda=\alpha^2 = \frac{b_0^2}{2\nu\Delta s}, \quad \mu = \frac{\alpha}{\beta} = \left|\frac{b_0}{b_1}\right|. \lb{eqA-10} \ee
See \cite{borodin2015handbook}, Part II, formula 2.02 (p.295). In the second form, this probability density is known as   
the {\it inverse Gaussian distribution} ${\rm IG}(\mu,\lambda)$ with parameters $\mu,$ $\lambda$ \citep{chhikara1988inverse}.  
A simple algorithm was devised by \cite{michael1976generating} to obtain realizations $\tilde{\sigma}_*$ drawn from the 
distribution ${\rm IG}(\mu,\lambda),$ by a transformation of a normal random variable $\tilde{N}$. One first 
constructs a chi-square random variable
\be \tilde{\chi} = \mu \tilde{N}^2 \lb{eqA-11} \ee   
and then sets
\be \tilde{\xi} = 1 + \left(\tilde{\chi} -\sqrt{4\lambda\tilde{\chi}+\tilde{\chi}^2}\right)/2\lambda.\lb{eqA-12} \ee
Finally, with probability $1/(1+\tilde{\xi})$ one selects $\tilde{\sigma}_*=\mu\tilde{\xi}$ and otherwise 
selects $\tilde{\sigma}_*=\mu/\tilde{\xi}.$ Below is a snippet of a Fortran code which implements 
this algorithm:  
{\small
$$
\begin{array}{l}
       \mbox{{\tt n1=normal}} \cr
       \mbox{{\tt chi=mu*n1**2}} \cr 
       \mbox{{\tt xi=chi-sqrt(abs(4*lambda*chi+chi**2))}} \cr 
       \mbox{{\tt xi=1+xi/(2*lambda)}} \cr 
       \mbox{{\tt u1=uniform}} \cr 
       \mbox{{\tt \textcolor{black}{if} (u1 .\textcolor{black}{le}. 1/(1+xi)) \textcolor{black}{then}}} \cr 
       \quad \mbox{{\tt     sigma=mu*xi}} \cr 
       \mbox{{\tt \textcolor{black}{else}}} \cr 
       \quad  \mbox{{\tt sigma=mu/xi}} \cr 
       \mbox{{\tt \textcolor{black}{end if}}} 
\end{array}
$$
}
For constant wall-normal velocity $v$ the above algorithm is exact with no constraint on the 
size of $\Delta s$ and it gives a statistically correct sampling of $\tilde{\sigma}_*$ and 
$\tilde{s}_*=\frac{s_0+\tilde{\sigma}_*s_1}{1+\tilde{\sigma}_*}.$ In an incompressible fluid flow the wall-normal 
velocity decreases to zero quadratically with the distance to the wall, so the algorithm should again provide 
an excellent approximation. 

Streamwise and spanwise locations of the stochastic Lagrangian particle at the first hitting time $\tilde{s}_*$
can be obtained in a statistically accurate fashion by a similar approach. For example, suppose that the 
streamwise position at the two integration times takes on the values $\tilde{a}(s_0)=a_0$, $\tilde{a}(s_1)=a_1.$ 
Between them it is accurately represented for small $\Delta s$ by the equation 
\be \tilde{a}(s) = a_0 + u(s-s_0) + \sqrt{2\nu}\, \tilde{W}_1(s), \lb{eqA-13} \ee
with $\tilde{W}_1(s)$ for $s<s_0$ another time-reversed Brownian motion satisfying $\tilde{W}_1(s_0)=0.$
Imposing the second end-point condition on $\tilde{a}(s_1)$ then implies that $\tilde{W}_1(s)$ becomes a generalized 
Brownian bridge satisfying 
\be \tilde{W}_1(s_0)=0, \quad \tilde{W}_1(s_1)= \frac{\Delta a - u\Delta s}{\sqrt{2\nu}}. \lb{eqA-14} \ee
We now make the same transformations as before, namely:  
\be \tilde{W}_1(s)= \left(\frac{\Delta a - u\Delta s}{\sqrt{2\nu}}\right) \frac{s-s_0}{s_1-s_0} + \tilde{B}^{(0)}_1(s)  \lb{eqA-15} \ee
for a standard Brownian bridge $\tilde{B}^{(0)}_1(s)$ and 
\be  \tilde{B}^{(0)}_1(s) = \frac{s_1-s}{\sqrt{\Delta s}} \tilde{B}_1(\sigma) = \frac{\sqrt{\Delta s}}{1+\sigma} \tilde{B}_1(\sigma) \lb{eqA-16} \ee
for a standard Brownian motion $\tilde{B}_1(\sigma)$ with $\sigma=\frac{s-s_0}{s_1-s}.$  Together these give
 \be   \tilde{a}(s) =  \frac{a_0 (s_1-s)+ a_1 (s-s_0)}{s_1-s_0} + \frac{\sqrt{2\nu\Delta s}}{1+\sigma} \tilde{B}_1(\sigma), \lb{eqA-17} \ee
 which is linear interpolation with a stochastic correction. In particular, the streamwise location at the time $\tilde{s}_*$ 
 of first hitting of the wall is 
\be   \tilde{a}_*:=\tilde{a}(\tilde{s}_*) =  \frac{a_0 (s_1-\tilde{s}_*)+ a_1 (\tilde{s}_*-s_0)}{s_1-s_0} 
+ \frac{\sqrt{2\nu\Delta s}}{1+\tilde{\sigma}_*} \tilde{B}_1(\tilde{\sigma}_*). \lb{eqA-18} \ee   
Likewise, the spanwise location is
\be   \tilde{c}_*:=\tilde{c}(\tilde{s}_*) =  \frac{c_0 (s_1-\tilde{s}_*)+ c_1 (\tilde{s}_*-s_0)}{s_1-s_0} 
+ \frac{\sqrt{2\nu\Delta s}}{1+\tilde{\sigma}_*} \tilde{B}_3(\tilde{\sigma}_*). \lb{eqA-19} \ee   
The two new Brownian motions at the single time $\tilde{\sigma}_*$ are statistically represented 
by $\tilde{B}_i(\tilde{\sigma}_*)=\sqrt{\tilde{\sigma}_*}\tilde{N}_i,$ $i=1,3$ where $\tilde{N}_1,$ $\tilde{N}_3$
are new independent normal random variables.

\bibliographystyle{jfm}

\bibliography{bibliography}

\providecommand{\noopsort}[1]{}\providecommand{\singleletter}[1]{#1}%
\begin{thebibliography}{72}
\expandafter\ifx\csname natexlab\endcsname\relax\def\natexlab#1{#1}\fi

\bibitem[Anderson(1966)]{anderson1966considerations}
{\sc Anderson, P.~W.} 1966 Considerations on the flow of superfluid helium.
  {\em Reviews of Modern Physics\/} {\bf 38}~(2), 298.

\bibitem[Arnold(1966)]{arnold1966geometrie}
{\sc Arnold, V.} 1966 Sur la g{\'e}om{\'e}trie diff{\'e}rentielle des groupes
  de {L}ie de dimension infinie et ses applications {\`a} l'hydrodynamique des
  fluides parfaits. {\em Annales de l'institut Fourier\/} {\bf 16}~(1),
  319--361.

\bibitem[Arouna(2004)]{arouna2004adaptative}
{\sc Arouna, Bouhari} 2004 Adaptative monte carlo method, a variance reduction
  technique. {\em Monte Carlo Methods and Applications\/} {\bf 10}~(1), 1--24.

\bibitem[Batchelor(2000)]{batchelor2000introduction}
{\sc Batchelor, G.K.} 2000 {\em An Introduction to Fluid Dynamics\/}. Cambridge
  University Press.

\bibitem[Berrut \& Trefethen(2004)]{berrut2004barycentric}
{\sc Berrut, Jean-Paul \& Trefethen, Lloyd~N} 2004 Barycentric lagrange
  interpolation. {\em SIAM review\/} {\bf 46}~(3), 501--517.

\bibitem[Besse \& Frisch(2017)]{besse2017geometric}
{\sc Besse, N. \& Frisch, U.} 2017 Geometric formulation of the {C}auchy
  invariants for incompressible {E}uler flow in flat and curved spaces. {\em
  Journal of Fluid Mechanics\/} {\bf 825}, 412--478.

\bibitem[Borodin \& Salminen(2015)]{borodin2015handbook}
{\sc Borodin, A.~N. \& Salminen, P.} 2015 {\em Handbook of Brownian Motion -
  Facts and Formulae\/}. Birkh{\"a}user Basel.

\bibitem[Box \& Muller(1958)]{box1958note}
{\sc Box, G.~E.~P. \& Muller, M.~E.} 1958 A note on the generation of random
  normal deviates. {\em The Annals of Mathematical Statistics\/} {\bf 29}~(2),
  610--611.

\bibitem[Boyer \& Fabrie(2012)]{boyer2012mathematical}
{\sc Boyer, F. \& Fabrie, P.} 2012 {\em Mathematical Tools for the Study of the
  Incompressible Navier-Stokes Equations and Related Models\/}. Springer, New
  York.

\bibitem[Brenier(2003)]{brenier2003topics}
{\sc Brenier, Yann} 2003 Topics on hydrodynamics and volume preserving maps. In
  {\em Handbook of mathematical fluid dynamics\/} (ed. S.~Friedlander \&
  D.~Serre), , vol.~2, pp. 55--86. Elsevier.

\bibitem[Brown \& Roshko(2012)]{brown2012turbulent}
{\sc Brown, Garry~L \& Roshko, Anatol} 2012 Turbulent shear layers and wakes.
  {\em Journal of Turbulence\/} {\bf 13}, N51.

\bibitem[Cauchy(1815)]{cauchy1815theorie}
{\sc Cauchy, A.~L.} 1815 Sur l'{\'{e}}tat du fluide {\`{a}} une {\'{e}}poque
  quelconque du mouvement. {M}{\'{e}}moires extraits des recueils de
  l'{A}cad{\'{e}}mie des sciences de l'{I}nstitut de {F}rance, {T}h{\'{e}}orie
  de la propagation des ondes \`{a} la surface d'un fluide pesant d'une
  profondeur ind{\'{e}}finie. {\em ({E}xtraits des {M}{\'{e}}moires
  pr{\'{e}}sent{\'{e}}s par divers savants \`{a} l’{A}cad{\'{e}}mie royale
  des {S}ciences de l'{I}nstitut de {F}rance et imprim{\'{e}}s par son ordre).
  {S}ciences math{\'{e}}matiques et physiques. {{\rm Tome I, 1827 Seconde
  Partie},}\/} pp. 33–--73.

\bibitem[Chhikara \& Folks(1988)]{chhikara1988inverse}
{\sc Chhikara, R. \& Folks, J.~L.} 1988 {\em The Inverse Gaussian Distribution:
  Theory: Methodology, and Applications\/}. Taylor \& Francis.

\bibitem[{Constantin} \& {Iyer}(2008)]{ConstantinIyer08}
{\sc {Constantin}, P. \& {Iyer}, G.} 2008 {A stochastic Lagrangian
  representation of the 3-dimensional incompressible Navier-Stokes equations}.
  {\em Commun. Pure Appl. Math.\/} {\bf 61}, 330--345.

\bibitem[{Constantin} \& {Iyer}(2011)]{ConstantinIyer11}
{\sc {Constantin}, P. \& {Iyer}, G.} 2011 {A stochastic-Lagrangian approach to
  the Navier–-Stokes equations in domains with boundary}. {\em The Annals of
  Applied Probability\/} {\bf 21.4}, 1466--1492.

\bibitem[Drivas \& Eyink(2017)]{drivas2017lagrangianII}
{\sc Drivas, Theodore~D \& Eyink, Gregory~L} 2017 A lagrangian
  fluctuation--dissipation relation for scalar turbulence. part ii.
  wall-bounded flows. {\em Journal of Fluid Mechanics\/} {\bf 829}, 236--279.

\bibitem[Eyink \& Spohn(1993)]{eyink1993negative}
{\sc Eyink, GL \& Spohn, H} 1993 Negative-temperature states and large-scale,
  long-lived vortices in two-dimensional turbulence. {\em Journal of
  statistical physics\/} {\bf 70}~(3-4), 833--886.

\bibitem[Eyink(2008)]{eyink2008turbulent}
{\sc Eyink, G.~L.} 2008 Turbulent flow in pipes and channels as cross-stream
  “inverse cascades” of vorticity. {\em Physics of Fluids\/} {\bf 20}~(12),
  125101.

\bibitem[Eyink(2010)]{eyink2010stochastic}
{\sc Eyink, G.~L.} 2010 Stochastic least-action principle for the
  incompressible {N}avier--{S}tokes equation. {\em Physica D: Nonlinear
  Phenomena\/} {\bf 239}~(14), 1236--1240.

\bibitem[Eyink {\em et~al.\/}(2019)Eyink, Gupta, Wang \& Zaki]{SCauchy2019}
{\sc Eyink, G.~L., Gupta, A., Wang, M. \& Zaki, T.} 2019 {{\tt SCauchy}} - {A
  {F}ortran90 code to calculate stochastic Cauchy invariants}.
  \url{https://github.com/mzwang2012/SCauchy.git}.

\bibitem[Eyink {\em et~al.\/}(2020{\natexlab{{\em a\/}}})Eyink, Gupta \&
  Zaki]{eyink2020stochasticII}
{\sc Eyink, G.~L., Gupta, A. \& Zaki, T.} 2020{\natexlab{{\em a\/}}} Stochastic
  lagrangian dynamics of vorticity. ii. channel-flow turbulence. J. Fluid Mech.
  (submitted).

\bibitem[Eyink {\em et~al.\/}(2020{\natexlab{{\em b\/}}})Eyink, Zaki \&
  Wang]{eyink2020stochastic}
{\sc Eyink, G.~L., Zaki, T. \& Wang, M.} 2020{\natexlab{{\em b\/}}} A
  stochastic {L}agrangian representation of the {L}ighthill-{M}orton theory. in
  preparation.

\bibitem[Falkovich {\em et~al.\/}(2001)Falkovich, Gawȩdzki \&
  Vergassola]{falkovich2001particles}
{\sc Falkovich, G, Gawȩdzki, K \& Vergassola, Massimo} 2001 Particles and
  fields in fluid turbulence. {\em Reviews of modern Physics\/} {\bf 73}~(4),
  913.

\bibitem[Freidlin(1985)]{freidlin1985functional}
{\sc Freidlin, M.~I.} 1985 {\em Functional Integration and Partial Differential
  Equations\/}. Princeton University Press.

\bibitem[Graham {\em et~al.\/}(2016)Graham, Kanov, Yang, Lee, Malaya, Lalescu,
  Burns, Eyink, Szalay, Moser \& Meneveau]{graham2016web}
{\sc Graham, J., Kanov, K., Yang, X.~I.~A., Lee, M., Malaya, N., Lalescu,
  C.~C., Burns, R., Eyink, G., Szalay, A., Moser, R.~D. \& Meneveau, C.} 2016 A
  web services accessible database of turbulent channel flow and its use for
  testing a new integral wall model for {LES}. {\em Journal of Turbulence\/}
  {\bf 17}~(2), 181--215.

\bibitem[Guala {\em et~al.\/}(2006)Guala, Liberzon, L{\"u}thi, Kinzelbach \&
  Tsinober]{guala2006stretching}
{\sc Guala, M., Liberzon, A., L{\"u}thi, B., Kinzelbach, W. \& Tsinober, A.}
  2006 Stretching and tilting of material lines in turbulence: the effect of
  strain and vorticity. {\em Physical Review E\/} {\bf 73}~(3), 036303.

\bibitem[Guala {\em et~al.\/}(2005)Guala, L{\"u}thi, Liberzon, Tsinober \&
  Kinzelbach]{guala2005evolution}
{\sc Guala, M., L{\"u}thi, B., Liberzon, A., Tsinober, A. \& Kinzelbach, W.}
  2005 On the evolution of material lines and vorticity in homogeneous
  turbulence. {\em Journal of Fluid Mechanics\/} {\bf 533}, 339--359.

\bibitem[Helmholtz(1858)]{helmholtz1858uber}
{\sc Helmholtz, H.~von} 1858 {\"U}ber {I}ntegrale der hydrodynamischen
  {G}leichungen welche den {W}irbelbewegungen entsprechen. {\em {J}ournal
  f{\"u}r die reine und angewandte {M}athematik\/} {\bf 55}, 25--55.

\bibitem[Higham {\em et~al.\/}(2013)Higham, Mao, Roj, Song \&
  Yin]{higham2013mean}
{\sc Higham, Desmond~J, Mao, Xuerong, Roj, Mikolaj, Song, Qingshuo \& Yin,
  George} 2013 Mean exit times and the multilevel monte carlo method. {\em
  SIAM/ASA Journal on Uncertainty Quantification\/} {\bf 1}~(1), 2--18.

\bibitem[Hofmann \& Math{\'e}(1997)]{hofmann1997quasi}
{\sc Hofmann, Norbert \& Math{\'e}, Peter} 1997 On quasi-monte carlo simulation
  of stochastic differential equations. {\em Mathematics of Computation\/} {\bf
  66}~(218), 573--589.

\bibitem[Huggins(1970)]{huggins1970energy}
{\sc Huggins, E.~R.} 1970 Energy-dissipation theorem and detailed {J}osephson
  equation for ideal incompressible fluids. {\em Physical Review A\/} {\bf
  1}~(2), 332.

\bibitem[Huggins(1994)]{huggins1994vortex}
{\sc Huggins, E.~R.} 1994 Vortex currents in turbulent superfluid and classical
  fluid channel flow, the {M}agnus effect, and {G}oldstone boson fields. {\em
  Journal of low temperature physics\/} {\bf 96}~(5-6), 317--346.

\bibitem[Johnson {\em et~al.\/}(2017)Johnson, Hamilton, Burns \&
  Meneveau]{johnson2017analysis}
{\sc Johnson, P.~L., Hamilton, S.~S., Burns, R. \& Meneveau, C.} 2017 Analysis
  of geometrical and statistical features of lagrangian stretching in turbulent
  channel flow using a database task-parallel particle tracking algorithm. {\em
  Physical Review Fluids\/} {\bf 2}~(1), 014605.

\bibitem[Johnson \& Meneveau(2016)]{johnson2016large}
{\sc Johnson, P.~L. \& Meneveau, C.} 2016 Large-deviation statistics of
  vorticity stretching in isotropic turbulence. {\em Physical Review E\/} {\bf
  93}~(3), 033118.

\bibitem[Josephson(1962)]{josephson1962possible}
{\sc Josephson, B.~D.} 1962 Possible new effects in superconductive tunnelling.
  {\em Physics letters\/} {\bf 1}~(7), 251--253.

\bibitem[Keanini(2006)]{keanini2006random}
{\sc Keanini, RG} 2006 Random walk methods for scalar transport problems
  subject to dirichlet, neumann and mixed boundary conditions. {\em Proceedings
  of the Royal Society A: Mathematical, Physical and Engineering Sciences\/}
  {\bf 463}~(2078), 435--460.

\bibitem[Kelvin(1868)]{kelvin1868vi}
{\sc Kelvin, L.} 1868 {VI}.--{O}n vortex motion. {\em Transactions of the Royal
  Society of Edinburgh\/} {\bf 25}~(1), 217--260.

\bibitem[Kim {\em et~al.\/}(1987)Kim, Moin \& Moser]{kim1987turbulence}
{\sc Kim, J., Moin, P. \& Moser, R.} 1987 Turbulence statistics in fully
  developed channel flow at low {R}eynolds number. {\em Journal of fluid
  mechanics\/} {\bf 177}, 133--166.

\bibitem[Kloeden \& Platen(2013)]{kloeden2013numerical}
{\sc Kloeden, P.~E. \& Platen, E.} 2013 {\em Numerical solution of stochastic
  differential equations\/}, , vol.~23. Springer Science \& Business Media.

\bibitem[Koumoutsakos(1999)]{koumoutsakos1999vorticity}
{\sc Koumoutsakos, P.} 1999 Vorticity flux control for a turbulent channel
  flow. {\em Physics of Fluids\/} {\bf 11}~(2), 248--250.

\bibitem[Kuz{'}min(1983)]{kuzmin1983ideal}
{\sc Kuz{'}min, G.~A.} 1983 Ideal incompressible hydrodynamics in terms of the
  vortex momentum density. {\em Physics Letters A\/} {\bf 96}~(2), 88--90.

\bibitem[Lee {\em et~al.\/}(2013)Lee, Malaya \& Moser]{lee2013petascale}
{\sc Lee, M., Malaya, N. \& Moser, R.~D.} 2013 Petascale direct numerical
  simulation of turbulent channel flow on up to 786{K} cores. In {\em SC '13,
  Proceedings of the International Conference on High Performance Computing,
  Networking, Storage and Analysis, Denver, Colorado — November 17 - 21,
  2013, {\rm {A}rticle 61}\/}. ACM, New York.

\bibitem[Li {\em et~al.\/}(2008)Li, Perlman, Wan, Yang, Meneveau, Burns, Chen,
  Szalay \& Eyink]{li2008public}
{\sc Li, Y., Perlman, E., Wan, M., Yang, Y., Meneveau, C., Burns, R., Chen, S.,
  Szalay, A. \& Eyink, G.} 2008 A public turbulence database cluster and
  applications to study {L}agrangian evolution of velocity increments in
  turbulence. {\em Journal of Turbulence\/} {\bf 9}, N31.

\bibitem[Lighthill(1963)]{lighthill1963boundary}
{\sc Lighthill, M.~J.} 1963 Boundary layer theory. In {\em Laminar Boundary
  Layers\/} (ed. L.~Rosenhead), pp. 46--113. Oxford University Press, Oxford.

\bibitem[L{\"u}thi {\em et~al.\/}(2005)L{\"u}thi, Tsinober \&
  Kinzelbach]{luthi2005lagrangian}
{\sc L{\"u}thi, B., Tsinober, A. \& Kinzelbach, W.} 2005 Lagrangian measurement
  of vorticity dynamics in turbulent flow. {\em Journal of Fluid mechanics\/}
  {\bf 528}, 87--118.

\bibitem[Lyman(1990)]{lyman1990vorticity}
{\sc Lyman, FA} 1990 Vorticity production at a solid boundary. {\em Appl. Mech.
  Rev\/} {\bf 43}~(8), 157--158.

\bibitem[Matsumoto(1997-present)]{matsumoto1997mersenne}
{\sc Matsumoto, M.} 1997-present . {M}ersenne {T}wister {H}ome {P}age.
  \url{http://www.math.sci.hiroshima-u.ac.jp/~m-mat/MT/emt.html}.

\bibitem[Matsumoto \& Nishimura(1998)]{matsumoto1998mersenne}
{\sc Matsumoto, M. \& Nishimura, T.} 1998 Mersenne twister: a 623-dimensionally
  equidistributed uniform pseudo-random number generator. {\em ACM Transactions
  on Modeling and Computer Simulation (TOMACS)\/} {\bf 8}~(1), 3--30.

\bibitem[Michael {\em et~al.\/}(1976)Michael, Schucany \&
  Haas]{michael1976generating}
{\sc Michael, J.~R., Schucany, W.~R. \& Haas, R.~W.} 1976 Generating random
  variates using transformations with multiple roots. {\em The American
  Statistician\/} {\bf 30}~(2), 88--90.

\bibitem[Morton(1984)]{morton1984generation}
{\sc Morton, B.~R.} 1984 The generation and decay of vorticity. {\em
  Geophysical \& Astrophysical Fluid Dynamics\/} {\bf 28}~(3-4), 277--308.

\bibitem[Oksendal(2013)]{oksendal2013stochastic}
{\sc Oksendal, Bernt} 2013 {\em Stochastic differential equations: an
  introduction with applications\/}. Springer Science \& Business Media.

\bibitem[Orszag(1971)]{orszag1971elimination}
{\sc Orszag, S.~A.} 1971 On the elimination of aliasing in finite-difference
  schemes by filtering high-wavenumber components. {\em Journal of the
  Atmospheric sciences\/} {\bf 28}~(6), 1074--1074.

\bibitem[Oseledets(1989)]{oseledets1989new}
{\sc Oseledets, V.~I.} 1989 On a new way of writing the {N}avier-{S}tokes
  equation. {T}he {H}amiltonian formalism. {\em Russian Mathematical Surveys\/}
  {\bf 44}~(3), 210.

\bibitem[Packard(1998)]{packard1998role}
{\sc Packard, R.~E.} 1998 The role of the {J}osephson-{A}nderson equation in
  superfluid helium. {\em Reviews of Modern Physics\/} {\bf 70}~(2), 641.

\bibitem[Panton(1984)]{panton1984incompressible}
{\sc Panton, R.L.} 1984 {\em Incompressible Flow\/}. John Wiley \& Sons.

\bibitem[Rapoport(2002)]{rapoport2002random}
{\sc Rapoport, Diego~L} 2002 Random diffeomorphisms and integration of the
  classical {N}avier-{S}tokes equations. {\em Reports on Mathematical
  Physics\/} {\bf 49}~(1), 1--27.

\bibitem[Rezakhanlou(2016)]{rezakhanlou2016stochastically}
{\sc Rezakhanlou, F.} 2016 Stochastically symplectic maps and their
  applications to the {N}avier--{S}tokes equation. {\em Annales de l'Institut
  Henri Poincare (C) Non Linear Analysis\/} {\bf 33}~(1), 1--22.

\bibitem[Roberts(1972)]{roberts1972hamiltonian}
{\sc Roberts, P.~H.} 1972 A {H}amiltonian theory for weakly interacting
  vortices. {\em Mathematika\/} {\bf 19}~(2), 169--179.

\bibitem[Salmon(1988)]{salmon1988hamiltonian}
{\sc Salmon, R.} 1988 Hamiltonian fluid mechanics. {\em Annual review of fluid
  mechanics\/} {\bf 20}~(1), 225--256.

\bibitem[Sawford(2001)]{sawford2001turbulent}
{\sc Sawford, Brian} 2001 Turbulent relative dispersion. {\em Annual review of
  fluid mechanics\/} {\bf 33}~(1), 289--317.

\bibitem[Shraiman \& Siggia(1994)]{shraiman1994lagrangian}
{\sc Shraiman, Boris~I \& Siggia, Eric~D} 1994 Lagrangian path integrals and
  fluctuations in random flow. {\em Physical Review E\/} {\bf 49}~(4), 2912.

\bibitem[Taylor(1932)]{taylor1932transport}
{\sc Taylor, G.~I.} 1932 The transport of vorticity and heat through fluids in
  turbulent motion. {\em Proceedings of the Royal Society of London. Series A,
  Containing Papers of a Mathematical and Physical Character\/} {\bf
  135}~(828), 685--702.

\bibitem[Taylor(1937)]{taylor1937statistical}
{\sc Taylor, G.~I.} 1937 The statistical theory of isotropic turbulence. {\em
  Journal of the Aeronautical Sciences\/} {\bf 4}~(8), 311--315.

\bibitem[Taylor(1938)]{taylor1938production}
{\sc Taylor, G.~I.} 1938 Production and dissipation of vorticity in a turbulent
  fluid. {\em Proceedings of the Royal Society of London. Series A-Mathematical
  and Physical Sciences\/} {\bf 164}~(916), 15--23.

\bibitem[Taylor \& Green(1937)]{taylor1937mechanism}
{\sc Taylor, G.~I. \& Green, A.~E.} 1937 Mechanism of the production of small
  eddies from large ones. {\em Proceedings of the Royal Society of London.
  Series A-Mathematical and Physical Sciences\/} {\bf 158}~(895), 499--521.

\bibitem[Tur \& Yanovsky(1993)]{tur1993invariants}
{\sc Tur, A.~V. \& Yanovsky, V.~V.} 1993 Invariants in dissipationless
  hydrodynamic media. {\em Journal of Fluid Mechanics\/} {\bf 248}, 67--106.

\bibitem[Varoquaux(2015)]{varoquaux2015anderson}
{\sc Varoquaux, Eric} 2015 Anderson's considerations on the flow of superfluid
  helium: Some offshoots. {\em Reviews of Modern Physics\/} {\bf 87}~(3), 803.

\bibitem[Weber(1868)]{weber1868ueber}
{\sc Weber, H.} 1868 {\"U}ber eine {T}ransformation der hydrodynamischen
  {G}leichungen. {\em Journal f{\"u}r die reine und angewandte Mathematik\/}
  {\bf 68}, 286--292.

\bibitem[Wu \& Wu(1993)]{wu1993interactions}
{\sc Wu, J-Z. \& Wu, J-M.} 1993 Interactions between a solid surface and a
  viscous compressible flow field. {\em Journal of Fluid Mechanics\/} {\bf
  254}, 183--211.

\bibitem[Wu \& Wu(1996)]{wu1996vorticity}
{\sc Wu, J-Z. \& Wu, J-M.} 1996 Vorticity dynamics on boundaries. {\em Advances
  in applied mechanics\/} {\bf 32}, 119--275.

\bibitem[Wu \& Wu(1998)]{wu1998boundary}
{\sc Wu, J-Z. \& Wu, J-M.} 1998 Boundary vorticity dynamics since {L}ighthill's
  1963 article: review and development. {\em Theoretical and computational
  fluid dynamics\/} {\bf 10}~(1-4), 459--474.

\bibitem[Zhao {\em et~al.\/}(2004)Zhao, Wu \& Luo]{zhao2004turbulent}
{\sc Zhao, H., Wu, J.-Z. \& Luo, J.-S.} 2004 Turbulent drag reduction by
  traveling wave of flexible wall. {\em Fluid Dynamics Research\/} {\bf
  34}~(3), 175.

\end{thebibliography}

\end{document}